\documentclass[12pt]{article}
\usepackage[latin9]{inputenc}
\usepackage{geometry}
\usepackage[]{graphicx}
\usepackage{epstopdf}
\usepackage{float}
\usepackage{mathtools}
\usepackage{amsmath}
\usepackage{amssymb}
\usepackage{graphicx}
\usepackage{setspace}
\usepackage{esint}
\usepackage[authoryear]{natbib}
\usepackage{appendix}
\input{undertilde}
\usepackage{bibunits}
\usepackage{titlesec}
\usepackage{algorithm,algorithmicx,algpseudocode,epsfig}
\usepackage{xr}
\makeatletter

\floatstyle{ruled}

\usepackage{latexsym}
\usepackage{amsthm}\usepackage{amsfonts}\usepackage{graphicx}\usepackage{epsfig}
\usepackage{bm}
\newcommand*{\patchAmsMathEnvironmentForLineno}[1]{%
      \expandafter\let\csname old#1\expandafter\endcsname\csname #1\endcsname
      \expandafter\let\csname oldend#1\expandafter\endcsname\csname end#1\endcsname
      \renewenvironment{#1}%
         {\linenomath\csname old#1\endcsname}%
         {\csname oldend#1\endcsname\endlinenomath}}%
    \newcommand*{\patchBothAmsMathEnvironmentsForLineno}[1]{%
      \patchAmsMathEnvironmentForLineno{#1}%
      \patchAmsMathEnvironmentForLineno{#1*}}%
    \AtBeginDocument{%
    \patchBothAmsMathEnvironmentsForLineno{equation}%
    \patchBothAmsMathEnvironmentsForLineno{align}%
    \patchBothAmsMathEnvironmentsForLineno{flalign}%
    \patchBothAmsMathEnvironmentsForLineno{alignat}%
    \patchBothAmsMathEnvironmentsForLineno{gather}%
    \patchBothAmsMathEnvironmentsForLineno{multline}%
    }
\usepackage[mathlines,displaymath]{lineno}

\include{def}
\def\dispmuskip{\thinmuskip= 3mu plus 0mu minus 2mu \medmuskip=  4mu plus 2mu minus 2mu \thickmuskip=5mu plus 5mu minus 2mu}
\def\textmuskip{\thinmuskip= 0mu                    \medmuskip=  1mu plus 1mu minus 1mu \thickmuskip=2mu plus 3mu minus 1mu}
\def\beq{\dispmuskip\begin{equation}}    \def\eeq{\end{equation}\textmuskip}
\def\beqn{\dispmuskip\begin{displaymath}}\def\eeqn{\end{displaymath}\textmuskip}
\def\bea{\dispmuskip\begin{eqnarray}}    \def\eea{\end{eqnarray}\textmuskip}
\def\bean{\dispmuskip\begin{eqnarray*}}  \def\eean{\end{eqnarray*}\textmuskip}

\newtheorem{theorem}{Theorem}
\newtheorem{lemma}[theorem]{Lemma}

\newtheorem{proposition}[theorem]{Proposition}

\newcommand{\wh}{\widehat}

\newcommand{\ov}{\overline}

\def\a{\alpha}

\def\d{\delta}

\def\N{{\cal N}}

\renewcommand{\d}{{\rm d}}

\def\utilde#1{\mathord{\vtop{\ialign{##\crcr
$\hfil\displaystyle{#1}\hfil$\crcr\noalign{\kern1.5pt\nointerlineskip}
$\hfil\tilde{}\hfil$\crcr\noalign{\kern1.5pt}}}}}
\def\undertilde#1{\mathord{\vtop{\ialign{##\crcr
$\hfil\displaystyle{#1}\hfil$\crcr\noalign{\kern1.5pt\nointerlineskip}
$\hfil\tilde{}\hfil$\crcr\noalign{\kern1.5pt}}}}}

\usepackage[mathlines,displaymath]{lineno}

\makeatother

\begin{document}
\title{Bayesian Inference for State Space Models using Block and Correlated Pseudo Marginal Methods}
\author{
P. Choppala\thanks{UNSW Business School, University of New South Wales}
\and D. Gunawan\footnotemark[1]
\and J. Chen \footnotemark[1]
\and M.-N. Tran\thanks{The University of Sydney Business School}
\and R. Kohn\footnotemark[1]
\thanks{The research of Choppala, Gunawan and Kohn was partially supported by the ARC Center of Excellence Grant CE140100049}
}
\maketitle

\begin{abstract}
This article addresses the problem of efficient Bayesian inference in dynamic systems using particle methods
and makes a number of contributions.
First, we develop a correlated pseudo-marginal (CPM) approach
for Bayesian inference in state space (SS) models that is based on filtering the disturbances, rather than the states. This approach is useful
when the state transition density is intractable or inefficient to compute, and also
when the dimension of the disturbance is lower than the dimension of the state.
Second, we propose a block pseudo-marginal (BPM) method that uses as the
estimate of the likelihood the average
of $G$ independent unbiased estimates of the likelihood.
We associate a set of underlying uniform of standard normal random numbers used to construct each of the individual unbiased likelihood
estimates and then use component-wise Markov Chain Monte Carlo to update the parameter vector jointly with one set of these random numbers at a time.
This induces a correlation of approximately $1-1/G$
between the logs of the estimated likelihood at the proposed and current values of the model parameters.
Third, we show for some non-stationary state space models
that the BPM approach is much more efficient than the CPM approach,
because  it is difficult to translate the high correlation in the underlying random numbers
to high  correlation between the logs of the likelihood estimates. Although our focus has been on applying the BPM method to state space models, our results and approach can be used in  a wide range of applications of the PM method, such as
panel data models, subsampling problems and approximate Bayesian computation.

Keywords: Averaged likelihood estimate, Correlated pseudo-marginal, Disturbance particle filter, Intractable Likelihood, Multiple processors
\end{abstract}

\section{Introduction}\label{sec:intro}
It is challenging to carry out Bayesian inference for the model parameters in SS models as
the likelihood is often intractable.
Because this likelihood at any given value of the parameters can be estimated unbiasedly using a particle filter (PF) \citep{Delmoral2004},
\cite{Andrieu2010a} propose using the pseudo-marginal (PM) Markov chain Monte Carlo (MCMC) based on an unbiased estimate of the likelihood to sample from the posterior distribution of the unknown model parameters.
 \cite{Pitt2012} show that to obtain a good tradeoff
 between computational complexity and MCMC mixing, the number of particles used in the PF should be such that the variance of the  log
 of the estimated likelihood (which we call $\sigma^2$)  is around one and that the inefficiency of the PM scheme increases exponentially with $\sigma^2$. Furthermore, the variance of the log of the estimated likelihood increases linearly with the sample size $T$ and is inversely proportional to the number of particles for a given sample size. Hence, the number of particles required to keep the variance around 1 is $O(T^2)$.

\cite{Dahlin:2015} and \cite{Deligiannidis2016} recently propose the CPM algorithm which correlates
 the log of the estimated likelihood at the proposed and current values of the model parameters by correlating
 the underlying standard normal random numbers used to construct the estimates of the likelihood.
 Introducing this correlation into the likelihood estimates reduces the variance of the
 difference in the logs of the estimated likelihoods which
 appears in the Metropolis-Hastings acceptance ratio.
\cite{Deligiannidis2016} show that the number of particles required by CPM in each MCMC iteration is $O(T^{3/2})$.
\cite{Minhgoc2016} propose an alternative PM approach, called the BPM approach,
which divides the uniform or standard normal random numbers into blocks
and then updates the unknown parameters jointly with one block of these random numbers, in each MCMC iteration.
\cite{Minhgoc2016} show that the number of particles required by BPM is $O(T^{3/2})$ if the likelihood is estimated using  Monte Carlo,
and is $O(T^{7/6})$ if the likelihood is estimated based on randomized quasi-Monte Carlo.

In some applications of SS models such as marine biogeochemical and economic models \citep{Murray2013,Jamie2014}, it is difficult
to estimate the likelihood unbiasedly using the standard particle filters that filter the state, because the state transition density is intractable or computationally expensive to compute.
\cite{Murray2013} express this intractable transition density using the disturbances in the state transition equation,
and reformulate the SS model by using the disturbances as the new state, who transition density is now tractable.
It is possible, and sometimes more efficient, to carry out a particle filter that is based on filtering these disturbances.
For example, in some SS models such as the Stochastic Volatility models in Section \ref{model:sim:sv4}, the disturbance dimension is lower than the state dimension.
We follow \cite{Murray2013} and refer to the SS model based on the disturbances as the Disturbance SS Model
 and the PF based on disturbances as the disturbance PF.

 The first contribution of this article is to extend the CPM for use with the disturbance PF.
The CPM algorithm operates on the space of states in such a way that the correlation in the likelihood values is preserved by maintaining similarity in the particle states. Our article
 proposes maintaining similarity in the disturbance space rather than in the state space.
 The motivating rationale is that if the disturbances are close to one another then the corresponding particle states will also be close to one another and hence the likelihood correlation will be preserved. Inducing strong positive correlation in the log of the estimated likelihood values
 leads to non-sticky mixing of the MCMC chain with fewer particles.
 In addition, it is more convenient and efficient to work with disturbances
 when their dimension is lower than that of the original states.

The CPM method involves a sorting procedure that can be computationally expensive; see Section \ref{sec:bayes_param_estim}.
The second contribution of this article is to propose a highly efficient BPM scheme for Bayesian inference in the general class of time series SS
models. The proposed BPM method extends the main idea in  \cite{Minhgoc2016} to time series models for which the unbiased likelihood estimator
is an average of likelihood estimates obtained by multiple independent PFs.
These PF's are run in parallel on multiple processors, with
the unknown model parameters and one block of the uniform or standard normal random numbers used in one of the PFs
updated jointly.
Using the power of multiple-processor architecture in the PM context has been explored
recently by \cite{Drovandi:2014}, but he did not incorporate the blocking idea.
This extension of the BPM approach in \cite{Minhgoc2016} leads to a surprisingly efficient Bayesian inference approach
for SS models.
 The theory in \cite{Minhgoc2016}
 makes it possible to obtain a desired correlation for the log of the
 estimated likelihoods to a sufficient accuracy, unlike the CPM method, and depends only on the number of independent PFs used.

Although our focus has been on applying the BPM method to SS models, our results and approach apply equally to a number of other applications of the PM method such as panel data models, subsampling problems and approximate Bayesian computation.

The rest of the paper is organised as follows: Section~\ref{sec:conv_ss_model} sets out the notation for the conventional SS model. Section~\ref{sec:conv_dis_model} presents the disturbance PF methodology. Section~\ref{sec:bayes_param_estim} considers the proposed CPM method for disturbance particle filtering.
Section~\ref{sec:blockPM} presents the BPM method. Section~\ref{sec:sim_study} illustrates the methodology using  simulated examples,
and section~\ref{sec:real_data} applies the methodology to real data. Section~\ref{sec:concl} concludes. There are two appendices. 
The first outlines how to estimate the gradient and hessian of the likelihood, and the second gives further details on the analysis of the simulated and real examples.

\section{Conventional SS model}\label{sec:conv_ss_model}
\begin{figure}[ht]
\centering
\includegraphics[width=.6\textwidth]{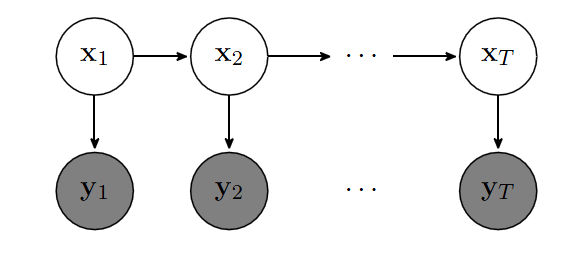}
\caption{Pictorial representation of the conventional SS model.}
\label{Fig:Conv_SSmodel}
\end{figure}

The SS model, shown in Fig.~\ref{Fig:Conv_SSmodel}, describes the evolution of a dynamic system with  $\mathbf{x}_{t}$ the state of the system at time $t$ and $\mathbf{y}_{t}$ the observation at time $t$.
The SS model can be mathematically described by  Markovian state transition densities and observation equation densities as follows
\begin{align}
\label{state_transit1}
\mathbf{x}_{t}|\mathbf{x}_{t-1} \sim g(\mathbf{x}_{t}|\mathbf{x}_{t-1}, \boldsymbol\theta),\; \; \mathbf{y}_{t}|\mathbf{x}_{t} \sim f(\mathbf{y}_{t}|\mathbf{x}_{t},\boldsymbol\theta), \; t=1,...,T
\end{align}
where, (a) the state vector is $\mathbf{x}_{t} \in \mathbb{R}^{n_{\mathbf{x}}}$ with $n_{\mathbf{x}}$ the state dimension, (b) the state transition density $g(\mathbf{x}_{t}|\mathbf{x}_{t-1}, \boldsymbol\theta)$ for $t\geq 2$ and $g(\mathbf{x}_{1}|\mathbf{x}_{0}, \boldsymbol\theta):=\mu(\mathbf{x}_{1}|\boldsymbol\theta)$ describes the evolution of the states, (c) the observation is $\mathbf{y}_{t} \in \mathbb{Y}^{n_{\mathbf{y}}}$ with $n_{\mathbf{y}}$ the dimension of the observations, (d) the observation density conditional on the state is
 $f(\mathbf{y}_{t}|\mathbf{x}_{t},\boldsymbol\theta)$, and (e) the model includes  a vector of unknown parameters $\boldsymbol\theta\in \Theta$. The observations $\mathbf{y}_{1:T}=\{\mathbf{y}_{1},\mathbf{y}_{2},...,\mathbf{y}_{T}\}$, corresponding to the states $\mathbf{x}_{1:T}=\{\mathbf{x}_{1},\mathbf{x}_{2},...,\mathbf{x}_{T}\}$, are the time series data that can be stock returns, electro-magnetic signals, camera images, etc.

State filtering and likelihood estimation 
 can be performed in a sequential Bayesian simulation framework 
 by approximating $p(\mathbf{x}_{1}|\mathbf{y}_{1},\boldsymbol\theta)$ and $p(\mathbf{y}_{1}|\boldsymbol\theta)$ at time $t=1$,
 $p(\mathbf{x}_{2}|\mathbf{y}_{1:2},\boldsymbol\theta)$ and $p(\mathbf{y}_{1:2}|\boldsymbol\theta)$ at time $t=2$, and so on.
 This is known as the PF in the literature.
 In the standard PF, it is necessary that we can generate from the state transition density $g(\mathbf{x}_{t}|\mathbf{x}_{t-1}, \boldsymbol\theta)$ and can evaluate
 the observation density $ f(\mathbf{y}_{t}|\mathbf{x}_{t},\boldsymbol\theta)$. Auxiliary particle filters often require that we can evaluate the state transition density $g(\mathbf{x}_{t}|\mathbf{x}_{t-1}, \boldsymbol\theta)$ as well as the observation density.

For the rest of the paper, we do not show dependence on $\boldsymbol\theta$, unless it is required.
 
\section{Disturbance SS model}\label{sec:conv_dis_model}
\begin{figure}[ht]
\centering
\includegraphics[width=.6\textwidth]{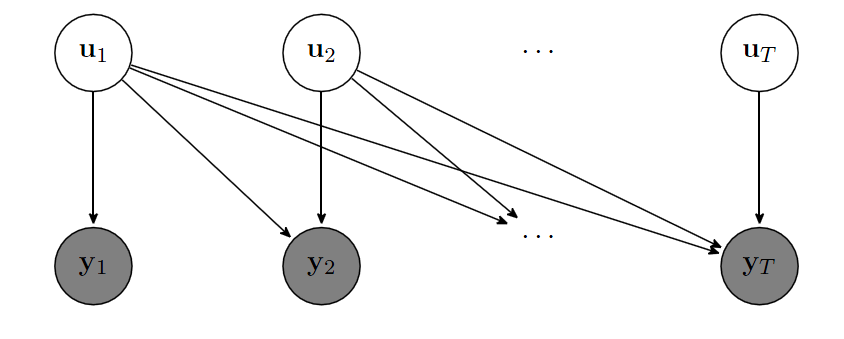}
\caption{Pictorial representation of the disturbance SS model.}
\label{Fig:Conv_Disturbmodel}
\end{figure}

The PF for the conventional SS model described in section~\ref{sec:conv_ss_model} assumes that
it is possible to either generate from or evaluate the state transition density $g(\mathbf{x}_{t}|\mathbf{x}_{t-1})$.
In some applications this is not possible or it is computationally expensive to compute \citep{Murray2013,Jamie2014}.
Suppose that, for $t\geq 2$, we can write $\mathbf{x}_{t}=k(\mathbf{u}_{t},\mathbf{x}_{t-1})$,
where $k(\cdot,\cdot)$ is a deterministic function and $\mathbf{u}_{t}\in\mathbb{R}^{n_{\mathbf{u}}}$ is a vector of latent variables with density $p_U(\mathbf u_t)$,
such that  $k(\mathbf{u}_{t},\mathbf{x}_{t-1})|\mathbf{x}_{t-1}\sim g(\mathbf{x}_{t}|\mathbf{x}_{t-1})$
when $\mathbf u_t\sim p_U(\mathbf u_t)$.
For $t=1$, $k(\mathbf{u}_{1},\mathbf{x}_{0}):=\kappa(\mathbf{u}_{1})$ with a deterministic function $\kappa(\cdot)$
such that  $\kappa(\mathbf{u}_{t})\sim \mu(\mathbf{x}_{1})$ when $\mathbf u_1\sim p_U(\mathbf u_1)$.
Typically, $\mathbf u_t$ is a set of independent uniform or standard normal random variables.
We follow \cite{Murray2013} and refer to the $\mathbf u_t$ as disturbances.
\cite{Murray2013} propose reformulating the SS model in terms of the disturbances variables $\mathbf{u}_{1:T}$
\begin{align}\label{state_transit2}
\mathbf{u}_{t} \sim p_U(\mathbf{u}_{t}),\; \; \mathbf{y}_{t}|\mathbf{u}_{1:t} \sim f(\mathbf{y}_{t}|\mathbf{u}_{1:t})=f(\mathbf{y}_{t}|\mathbf{x}_{t}), \; t=1,...,T,
\end{align}
where $\mathbf{x}_{t}=\mathbf{x}_{t}(\mathbf u_{1:t})=k(\mathbf{u}_{t},\mathbf{x}_{t-1})$.
\cite{Murray2013} call this the disturbance SS model.  Figure~\ref{Fig:Conv_Disturbmodel} gives its graphical dependence.
It is possible to use a PF for the disturbance SS model \eqref{state_transit2} as the state transition density $p(\mathbf{u}_{t}|\mathbf{u}_{t-1})=p_U(\mathbf{u}_{t})$ is tractable.
We note that, unlike the conventional SS model \eqref{state_transit1},
the conditional distribution of $\mathbf{y}_{t}$ in \eqref{state_transit2} depends on all the state disturbances $\mathbf u_{1:t}$  up to time $t$.
State filtering and likelihood estimation proceeds by estimating $p(\mathbf{u}_{1}|\mathbf{y}_{1})$ and $p(\mathbf{y}_{1})$ at time $t=1$,
 $p(\mathbf{u}_{1:2}|\mathbf{y}_{1:2})$ and $p(\mathbf{y}_{1:2})$ at time $t=2$, etc.
The likelihood in the conventional SS model \eqref{state_transit1} is
\begin{align}\label{likelihood1}
p(\mathbf{y}_{1:T}) = \int \bigg[ \prod_{t=1}^{T} f(\mathbf{y}_{t}|\mathbf{x}_{t}) g(\mathbf{x}_{t}|\mathbf{x}_{t-1}) \bigg]\d \mathbf x_{1:T},
\end{align}
and the likelihood in the disturbance SS model \eqref{state_transit2} is
\begin{align}\label{likelihood2}
p(\mathbf{y}_{1:T}) =\int \bigg[ \prod_{i=1}^{T} f\Big(\mathbf{y}_{t}|k(\mathbf{u}_{t},\mathbf{x}_{t-1})\Big) p(\mathbf{u}_{t}) \bigg]\d \mathbf u_{1:T}.
\end{align}
Proposition~\ref{proposition1} shows that \eqref{likelihood1} and \eqref{likelihood2} are equal. Its proof is obvious and omitted.
\begin{proposition}\label{proposition1}
Suppose that for each value of $\mathbf{x}_{t-1}$, the mapping $\mathbf{u_t}\mapsto \mathbf x_t=k(\mathbf{u}_{t},\mathbf{x}_{t-1})$
from $\mathbb{R}^{n_{\mathbf{u}}}$ into $\mathbb{R}^{n_{\mathbf{x}}}$ is one-to-one.
Then \eqref{likelihood1} and \eqref{likelihood2} are equal.
\end{proposition}
Therefore we can carry out Bayesian inference for the model parameters $\boldsymbol\theta$ using the PM approach,
where the likelihood \eqref{likelihood2} is estimated unbiasedly based on a PF operating on the disturbance SS model \eqref{state_transit2}.
The next section presents the PF methodology for the disturbance SS model.

\subsection{Particle filter for the disturbance SS model} \label{sec:dpf}
Sequential particle filtering for the disturbance SS model involves
approximating $p(\mathbf{u}_{1:t}|\mathbf{y}_{1:t})$
and the likelihood $L_t=\int p(\mathbf{u}_{1:t},\mathbf{y}_{1:t})\d\mathbf{u}_{1:t}$,
based on a set of weighted samples $\{\mathbf{u}_{1:t}^i,W_t^i\}_{i=1}^N$, $t=1,2,..$.
The conventional state can be recovered at every time step by $\mathbf x_t^i=k(\mathbf u_t^i,\mathbf x_{t-1}^i)$.
The PF scheme is basically a sequential importance sampling and resampling procedure
with the proposal density at step $t$ having the following structure
\[q_t(\mathbf{u}_{1:t})=q_{t-1}(\mathbf{u}_{1:t-1})q_{t}(\mathbf{u_t}|\mathbf{u}_{1:t-1}).\]
Note that $q_{t}(\mathbf{u_t}|\mathbf{u}_{1:t-1})$ might depend on the data $\mathbf{y}_{1:t}$ as in the fully adapted PF.
The unnomarlized weights $w_t(\mathbf{u}_{1:t})={p(\mathbf{u}_{1:t},\mathbf{y}_{1:t})}/{q_t(\mathbf{u}_{1:t})}$ are decomposed as
\[w_t(\mathbf{u}_{1:t})=w_{t-1}(\mathbf{u}_{1:t-1})\a_{t}(\mathbf{u}_{1:t}),\;\;\text{with}\;\;
\a_{t}(\mathbf{u}_{1:t})=\frac{f(\mathbf{y}_t|\mathbf{u}_{1:t})p_U(\mathbf u_t)}{q_{t}(\mathbf{u_t}|\mathbf{u}_{1:t-1})}=\frac{f(\mathbf{y}_t|\mathbf{x}_{t})p_U(\mathbf u_t)}{q_{t}(\mathbf{u_t}|\mathbf{u}_{1:t-1})}.\]
It is easy to show that
\[\frac{L_t}{L_{t-1}}=\frac{\int \alpha_t(\mathbf{u}_{1:t})w_{t-1}(\mathbf{u}_{1:t-1})q_t(\mathbf{u}_{1:t})\d\mathbf{u}_{1:t}}{\int w_{t-1}(\mathbf{u}_{1:t-1})q_t(\mathbf{u}_{1:t})\d\mathbf{u}_{1:t}},\]
which can be estimated by
\begin{equation}\label{llh_ratio}
\wh{\frac{L_t}{L_{t-1}}}=\sum_{i=1}^NW_{t-1}^i\alpha_t(\mathbf{u}_{1:t}^i)
\end{equation}
with $W_{t-1}^i$ the normalized weights from time $t-1$.
Therefore the product of the estimates in \eqref{llh_ratio} up to time $t$ forms an estimate $\wh L_t$ of the likelihood $L_t$.
By definition, $L_0=1$. It is well-known that $\wh L_t$ is an unbiased estimate of $L_t$ \citep{Delmoral2004}.
Algorithm~\ref{algo_dis_sir} gives the pseudo-code of the disturbance PF.

\begin{algorithm}
\caption{$[\{\mathbf{x}_{t}^{i},W_{t}^{i}\}_{i=1}^{N},\wh L_t]$=PF}
\label{algo_dis_sir}
\begin{algorithmic}
\item At time $t=1$
\begin{algorithmic}
\item sample $\mathbf{u}_{1}^{i} \sim q_1(\mathbf{u}_{1}|\mathbf{y}_{1})$, $i=1,...,N$
\item set $\mathbf{x}_{1}^{i}=\kappa(\mathbf{u}_{1}^{i})$, $i=1,...,N$
\item compute unnormalized weights $w_{1}^{i} = \alpha(\mathbf u_1^i)$, and normalized weights $W_{1}^{i}\propto w_{1}^{i}$
\item $\wh L_1\longleftarrow\frac{1}{N}\sum \alpha(\mathbf u_1^i)$
\item If {$\big( \sum_{i=1}^{N} (W_{1}^{i})^{2} \big)^{-1}<N_{\text{threshold}}$}
\begin{algorithmic}
\item       $[\{I(j)\}_{j=1}^{N}]=\text{Resample}[\{W_{1}^{i}\}_{i=1}^{N}]$
\item       For $i=1:N$, assign particles $\mathbf{u}_{1}^{i} \longleftarrow \mathbf{u}_{1}^{I(i)}  $ and reset weights $W_{1}^{i} = 1/N$
\end{algorithmic}
\item  \}
\end{algorithmic}
\item At time $t\geq2$
\begin{algorithmic}
\item sample $\mathbf{u}_{t}^{i} \sim q_t(\mathbf{u}_{t}|\mathbf{u}_{1:t-1}^{i},\mathbf{y}_{t})$, $i=1,...,N$
\item set $\mathbf{x}_{t}^{i}=k(\mathbf{u}_{t}^{i},\mathbf{x}_{t-1}^{i})$, $i=1,...,N$
\item compute weights $w_{t}^{i} = w_{t-1}^{i}\times \alpha(\mathbf u_{1:t}^i)$, and $W_{t}^{i}\propto w_{t}^{i}$, $i=1,...,N$
\item $\wh L_t=\wh L_{t-1}\times \sum W_{t-1}^i\alpha(\mathbf u_{1:t}^i)$
\item If {$\big( \sum_{i=1}^{N} (W_{t}^{i})^{2} \big)^{-1}<N_{\text{threshold}}$}
\begin{algorithmic}
\item       $[\{I(j)\}_{j=1}^{N}]=\text{Resample}[\{W_{t}^{i}\}_{i=1}^{N}]$
\item       For $i=1:N$, assign particles $\mathbf{u}_{1:t}^{i} \longleftarrow \mathbf{u}_{1:t}^{I(i)}  $ and reset weights $W_{t}^{i} = 1/N$
\end{algorithmic}
\item  \}
\end{algorithmic}
\end{algorithmic}
\end{algorithm}

\section{CPM for the disturbance SS model}\label{sec:bayes_param_estim}
The PF provides an unbiased estimate of the likelihood and this property facilitates the development of estimation techniques using the PM as in \cite{Andrieu2010a}, \cite{Murray2013} and \cite{Jamie2014}. This section proposes a CPM algorithm for Bayesian inference in the disturbance SS model.

The unbiased non-negative PF likelihood estimator can be written as $\wh L(\boldsymbol\theta,\mathbf{U})$, a function of the model parameters and the collection $\mathbf{U}$ of all the standard normal random numbers (or equivalently uniform random numbers)  used to obtain the likelihood estimate.
The collection $\mathbf{U}$ consists of, in the conventional PF, the $N$ $n_{\mathbf{x}}-$dimensional random vectors used to generate the new states $\mathbf x_t$ at each time step $t=1,...,T$, and in the disturbance PF, the $N$ $n_{\mathbf{u}}-$dimensional random vectors used to generate $\mathbf u_t$ at each time step $t=1,...,T$.
Moreover resampling involves additional random numbers. This paper employs systematic resampling which requires one random number per resampling step.
Denote by $p_U(\mathbf{U})$ the density of $\mathbf{U}$.
The CPM method is derived as follows. Define the joint pseudo target of $\boldsymbol\theta$ and $\mathbf{U}$ as
\begin{align}
\label{pseudo_target1}
\ov{\pi}(\boldsymbol\theta,\mathbf{U}) = \cfrac{\wh L(\boldsymbol\theta,\mathbf{U}) \; p_U(\mathbf{U}) \; p_\Theta(\boldsymbol\theta)}{p(\mathbf{y}_{1:T})}.
\end{align}
The marginal $\ov \pi(\theta)$ of $\ov{\pi}(\boldsymbol\theta,\mathbf{U})$ with respect to $\boldsymbol\theta$ is the same as the posterior $\pi(\boldsymbol\theta)=p(\mathbf{y}_{1:T}|\boldsymbol\theta)p_\Theta(\boldsymbol\theta)/p(\mathbf{y}_{1:T})$, because $\int \wh L(\boldsymbol\theta,\mathbf{U}) \; p_U(\mathbf{U})\d\mathbf{U}=p(\mathbf{y}_{1:T}|\boldsymbol\theta)$.
That is, we can run an MCMC on the expanded space of $(\boldsymbol\theta, \mathbf{U})$, and obtain iterates
 from  $\pi(\boldsymbol\theta)$. Let $\boldsymbol\theta^{c}$ and $\mathbf{U}^{c}$ be the current values of the parameters and the random numbers respectively. To iterate through the MCMC chain, we first propose their corresponding new values according to
\begin{align}
\boldsymbol\theta^{p} \sim q(\boldsymbol\theta^{p}|\boldsymbol\theta^{c})\quad \text{and} \quad
\mathbf{U}^{p} \sim q(\mathbf{U}^{p}|\mathbf{U}^{c})=p_U(\mathbf{U}^{p})
\end{align}
and then accept this proposal with the probability
\begin{align}
\alpha &= \text{min} \bigg(1, \; \cfrac{\wh L(\boldsymbol\theta^{p},\mathbf{U}^{p}) \; p_\Theta(\boldsymbol\theta^{p}) \; p_U(\mathbf{U}^{p})}{\wh L(\boldsymbol\theta^{c},\mathbf{U}^{c}) \; p_\Theta(\boldsymbol\theta^{c}) \; p_U(\mathbf{U}^{c})} \times \cfrac{q(\boldsymbol\theta^{c}|\boldsymbol\theta^{p}) \; p_U(\mathbf{U}^{c})}{q(\boldsymbol\theta^{p}|\boldsymbol\theta^{c}) \; q_U(\mathbf{U}^{p})} \bigg) \nonumber \\
&= \text{min} \bigg(1, \; \cfrac{\wh L(\boldsymbol\theta^{p},\mathbf{U}^{p}) \; p_\Theta(\boldsymbol\theta^{p})}{\wh L(\boldsymbol\theta^{c},\mathbf{U}^{c}) \; p_\Theta(\boldsymbol\theta^{c})} \times \cfrac{q(\boldsymbol\theta^{c}|\boldsymbol\theta^{p}) }{q(\boldsymbol\theta^{p}|\boldsymbol\theta^{c})} \bigg).  \label{pmmh_accept_prob}
\end{align}
This standard PM method has been applied to both the conventional and the disturbance SS models \citep{Andrieu2010a,Pitt2012,Murray2013}.

An important version of the PM algorithm is the CPM algorithm of \cite{Dahlin:2015} and \cite{Deligiannidis2016}
 where the likelihood estimates appearing in the MH ratio
 are correlated by correlating the current and proposed standard normal random numbers $\mathbf{U} $  as
\begin{align}
\label{corr_rand}
\mathbf{U}^{p} = \rho \; \mathbf{U}^{c} + \sqrt{1-\rho^2} \; {\mathbf{\xi}}
\end{align}
where $ \rho$ is the non-negative correlation between the random numbers and ${\mathbf{\xi}}$ is a standard normal vector of the same size as $\mathbf U$.
The standard PM is the special case of CPM when $\rho=0$. The CPM algorithm accelerates the estimation process significantly. The strong correlation between logs of the likelihood
estimates  reduces the variability in the likelihood ratio of the proposed and current parameter values resulting in faster convergence.

In this paper, we propose using CPM for the disturbance SS model.
It is important that the correlation in the random numbers is preserved in the estimated likelihoods. However the resampling step of the PF might impede this preservation because of its particle replacement property \citep{Deligiannidis2016}. A small change in the random numbers used in the resampling steps
might lead to a big change in the particle paths, and thus the correlation in the logs of the likelihood estimates might not be well preserved.
This impediment can be facilitated in two steps.

\noindent{\bf Fixed resampling frequency:} Resampling is usually performed whenever the effective sample size $N_{\text{eff}}$ falls below a threshold. In our CPM the particles are resampled once every $R_f$ time steps with $1 \leq R_f \leq T$. Having a deterministically fixed resampling frequency $R_f$ allows particles to be prearranged in a way that preserves the correlation.

\noindent
{\bf Particle ordering:} Prior to being resampled, the particles are ordered so that they are close to one another in some metric. Ordering can be easily accomplished in the univariate case by sorting the particles from smallest to the largest. For the multivariate case, such sorting is unavailable. One approach for ordering multidimensional particles is to use a Hilbert space filling curve method \citep{Skilling2004}, which is used in the PF by
\cite{Gerber2015} (in a context not related to CPM), and is proposed for the CPM algorithm by \cite{Deligiannidis2016}. The Hilbert curve method  transforms multidimensional particles to a univariate space (the Hilbert space in this context) based on some metric so that particle locality is preserved. The resulting transformed univariate set of particles is then sorted to obtain the sorting indices, i.e.,
the Hilbert curve method provides a mechanism to transform multidimensional particles to a space on which traditional sorting can be applied.
Our article proposes the following simpler and more resource-efficient multidimensional sorting scheme.

\subsection{Multidimensional Euclidean sorting}
Let $\{\mathbf{x}^{i}\}_{i=1,...,N}$ be the $n_{\mathbf{x}}-$dimensional particles at a given time step, $\mathbf{x}^{i} = (x_{1}^{i},...,x_{n_{\mathbf{x}}}^{i})^{\top}$. The time subscript is removed for notational simplicity.
Let $d(\mathbf{x}^{j},\mathbf{x}^{i})$ be the Euclidean distance between two multidimensional particles $\mathbf{x}^{i}$ and $\mathbf{x}^{j}$.
Algorithm~\ref{euclid_sort} describes the procedure to generate the set of sorting indices for the particles.
The first sorting index in the algorithm is the index of the particle having the least value along its first dimension.
The rest are chosen in a way that minimizes the Euclidean distance between the recently selected particle and the set of all remaining particles. This approach is employed in our article to sort multidimensional particles.
\begin{algorithm}
\caption{$S$=Euclidean-Sorting$[\{\mathbf{x}^{i}\}_{i=1,...,N}]$}
\label{euclid_sort}
\begin{algorithmic}
\item FOR \{ {$ j=1$}
\item Form index set $\chi^{j} = \{1,...,N\}$
\item Obtain sorting index $S(j)=\text{min}_{i} \; x_{1}^{i} \; \forall \; i \; \in \; \chi^{j}$
\item \}
\item FOR \{{$ j=2,...,N$}
\item Set particle $\mathbf{x}^{*} \longleftarrow \mathbf{x}^{j-1}$
\item Update index set $\chi^{j} = \chi^{j-1}\setminus S(j-1)$
\item Obtain sorting index $S(j)=\text{min}_{i} \; d(\mathbf{x}^{*},\mathbf{x}^{i}) \; \forall \; i \; \in \; \chi^{j}$
\item \}
\end{algorithmic}
\end{algorithm}

The two PF modifications -- fixed resampling frequency and particle ordering -- ensure that the variation in the particle states due to resampling is minimized. Hence the correlation in the likelihood is more likely to be preserved leading to good mixing of the MCMC chain with fewer particles.

\subsection{CPM for the disturbance SS models}
The CPM can be extended to the disturbance SS models  by ordering the original states. However, we take a different approach by working in the disturbance space. The justification is that, since the particles are often a smooth function of the disturbances, a similarity (or variation) in the disturbances leads to similarity (or variation) in the state particle. If the disturbances  are close to each other
then the corresponding state particles  will also be close to each other. This implies that sorting the disturbances is sufficient to preserve the correlation in the PF likelihood values. The key advantages of this proposal are, (a) the strong positive correlation in the likelihood values leads to faster convergence of the MCMC chain with fewer particles, and (b) further acceleration is possible when disturbances have  lower dimension than the  states. Performing filtering and estimation directly on the lower dimensional disturbance space facilitates a further reduction in the number of particles and hence in the computational complexity.

\section{BPM for SS models}
\label{sec:blockPM}
The CPM method involves particle sorting in order to maintain proximity within the particles. This procedure can be computationally expensive and renders the CPM method infeasible for high dimensional models involving a large number of observations.
Furthermore, the correlation in the log likelihood estimates
is controlled by the correlation $\rho$ between the $\mathbf U$,
but this relation is model-dependent and not known precisely.
A large $\rho$ does not necessarily lead to a high correlation in the log likelihood estimates
(see section~\ref{sec:sim_study}).
These limitations can be overcome in the BPM approach proposed in this section.

The BPM method of \cite{Minhgoc2016} is an alternative to the CPM,
in which the set $\mathbf U$ is divided into $G$ blocks, each of these is updated jointly with $\boldsymbol{\theta}$ in each MCMC iteration.
Under some assumptions, they show that the optimal number of particles should be selected such that the variance of the log-likelihood estimate is $\sigma_\text{opt}^2=2.16^2/(1-\rho^2_l)$
with $\rho_l$ the correlation between the log of the likelihood estimates. We extend the BPM  to the general class of SS models where the unbiased likelihood estimate is an average of $G$ likelihood estimates
obtained by $G$ independent PFs.

Let $\widehat{L}\left(\boldsymbol{\theta},\mathbf{U}_{\left(i\right)}\right)$ be the estimated likelihood obtained from the $i$th PF, $i=1,...,G$. We define the joint target density of $\boldsymbol\theta$ and $\mathbf{U}=
\left ( \mathbf{U}_{\left(1\right)}, \dots ,  \mathbf{U}_{\left(G\right)}\right ) $  as
\begin{align}
\ov \pi\left(\boldsymbol{\theta},\mathbf{U}\right) & = \ov{\widehat{L}}\left(\boldsymbol{\theta},\mathbf{U}\right)
p_\Theta \left(\boldsymbol{\theta}\right)\prod_{i=1}^G
p_{U}\left(\mathbf{U}_{\left(i\right)}\right)/p\left (\mathbf{y}_{1:T} \right)\\
\intertext{where}
\ov{\widehat{L}}\left(\boldsymbol{\theta},\mathbf{U}\right)&:=\frac1G\sum_{i=1}^G\widehat{L}\left(\boldsymbol{\theta},
\mathbf{U}_{\left(i\right)}\right)
\end{align}
is the average of the $G$ unbiased likelihood estimates and hence also unbiased.
We then update the parameters jointly with a randomly-selected block $\mathbf{U}_{\left(K\right)}$ in each MCMC iteration,
with $\Pr\left(K=k\right)=1/G$ for any $k=1,...,G$. Using this scheme, the acceptance probability is
\begin{equation}
\alpha = \min\left\{ 1,\frac{\ov{\widehat{L}}\left(\boldsymbol{\theta}^{p},\mathbf{U}_{\left(1\right)}^{c},...,\mathbf{U}_{\left(k-1\right)}^{c},\mathbf{U}_{\left(k\right)}^{p},\mathbf{U}_{\left(k+1\right)}^{c},...,\mathbf{U}_{\left(G\right)}^{c}\right)p_\Theta\left(\boldsymbol{\theta}^{p}\right)}{\ov{\widehat{L}}\left(\boldsymbol{\theta}^{c},\mathbf{U}_{\left(1\right)}^{c},...,\mathbf{U}_{\left(k-1\right)}^{c},\mathbf{U}_{\left(k\right)}^{c},\mathbf{U}_{\left(k+1\right)}^{c},...,\mathbf{U}_{\left(G\right)}^{c}\right)p_\Theta\left(\boldsymbol{\theta}^{c}\right)}\frac{q\left(\boldsymbol{\theta}^{c}|\boldsymbol{\theta}^{p}\right)}{q\left(\boldsymbol{\theta}^{p}|\boldsymbol{\theta}^{c}\right)}\right\}.
\label{eq:acceptance prob}
\end{equation}
The PFs can be run in parallel on multiple processors. It is possible to show that the correlation between $\log\widehat{L}\left(\boldsymbol{\theta}^{p},\mathbf{U}^{p}\right)$ and $\log\widehat{L}\left(\boldsymbol{\theta}^{c},\mathbf{U}^{c}\right)$ is approximately $\rho_l=1-1/G$, so the more particle filters we run the higher the correlation.
Unlike the CPM, the BPM method allows a more direct control of the correlation between the logs of the estimated likelihoods, and thus provides
a principled way to select the number of particles in each PF.

\section{Simulation studies}\label{sec:sim_study}
This section presents the simulation studies for the proposed PM methods in a wide range of models including
 (a)  a stochastic volatility model, (b) a non-stationary growth model, (c) a spline model, and (d) the Lotka Volterra model.
 A commonly used performance measure in MCMC is the integrated autocorrelation time (IACT).
 For a univariate parameter $\theta$, IACT is estimated by
\begin{align}
\text{IACT} := 1+2\sum_{t=1}^{1000} \wh\rho_t
\end{align}
where $\wh\rho_t$ are the sample autocorrelations.
For a multivariate $\boldsymbol\theta$, we report the average $\overline{\text{IACT}}$ of IACTs over the coordinates.
The efficiency of a sampling scheme becomes evident when the MCMC performance is studied jointly with the time taken to run the simulation, which  is measured by the time normalized variance (TNV) defined as
\begin{align}
\text{TNV} := {\text{IACT}} \times \text{Time},
\end{align}
where Time is the elapsed time (in seconds) per iteration. 
We also compute the relative time normalized variance (RTNV) defined as
\begin{align}
\text{RTNV} = \text{TNV}/\text{TNV}_{\text{benchmark}},
\end{align}
where the benchmark is chosen to be the standard PM method for all the models studied in this paper.

In all the studies, the total number of MCMC iterations is $25000$, with the first $5000$ discarded as burnin iterations.
The particles are resampled at every time step.
The adaptive random walk (ARW) algorithm of \cite{GarethRoberts2009} is used,
with the  scale in the ARW proposal  adapted using the Robbins-Monro method of \cite{ScottSisson2016} to obtain an acceptance rate of $0.23$.
The proposal using the gradient information from the likelihood is also used for the BPM method.
See the Appendix for the details.
The number of particles in the standard PM method is chosen such that $ \text{Var}(\log\wh L(\boldsymbol\theta)) \approx 1$. The number of particles in the CPM methods (with particle and disturbance sorting) is chosen such that $\text{Var}(\log\wh L(\boldsymbol\theta))\approx 2.16^{2}/\left(1-\rho^{2}_l\right)$ where $\rho_l$ is the log likelihood correlation estimated over $50$ independent replications of the particle filter with the parameters fixed at their true values.
For the BPM method, we use $G=12$ PFs, hence $\rho_l=1-1/12=0.9167$ and the number of particles is chosen such that $ \text{Var}(\log\wh L(\boldsymbol\theta)) \approx 29.21$.

\subsection{Stochastic Volatility model}\label{model:sim:sv4}
The stochastic volatility (SV) model is composed of a linear state transition equation and a nonlinear (in the states)
observation equation. The state transition model is a $P$th order linear Gaussian autoregressive process  $\text{AR}(P)$  with $P \geq 1$
\begin{align}
 v_{t} = \sum_{i=1}^{P} \phi_{i} v_{t-i} + a_{t}, \; a_{t} \sim \mathcal{N}(0,\tau^2)
\end{align}
The state transition and the observation equations of the corresponding SS model are
\begin{gather}
\mathbf{x}_{t} = \mathbf{F} \; \mathbf{x}_{t-1} + \mathbf{C} \; a_{t} \\
y_{t} = \text{exp}(x_{1,t} / 2) \; e_{t},\;\;e_{t} \sim \mathcal{N}(0,1)
\end{gather}
where 
\[
\underset{(P \times P)}{\mathbf{F}}=
  \begin{pmatrix}
    \phi_{1} & \phi_{2} &. & . & . & \phi_{P-1} & \phi_{P} \\
     1 & 0 & . & . & . & 0 & 0 \\
     0& 1& .&.&.& 0& 0 \\
     .  \\
     .  \\
    0& 0& .&.&.& 1& 0
  \end{pmatrix}, \;
  \underset{(P \times 1)}{\mathbf{C}}=
  \begin{pmatrix}
    1 \\
    0 \\
    0 \\
    .  \\
    .  \\
    0
  \end{pmatrix}.
\]
The transition equation can be written as $\mathbf{x}_{t} = k(\mathbf{x}_{t-1},u_t)= \mathbf{F} \; \mathbf{x}_{t-1} + \tau\mathbf{C} \; u_{t} $, $u_t\sim \N(0,1)$.
The original state dimensionality is $n_{\mathbf{x}}= P$ and the disturbance dimensionality is $n_{\mathbf{u}}=1$, hence when $P>1$ the disturbance has a lower dimension than the state.

We chose $\phi_{i}=\phi/P$ for $i=1,...,P$. The parameters for this model are $\boldsymbol\theta=(\phi, \tau^2)$.
The true parameter values are $\phi_{\text{true}}=0.98$ and $\tau^2_{\text{true}}=0.1$.
The priors for the parameters are $\text{log } \phi \sim \mathcal{N}(0,50)$ and $\text{log } \tau^2 \sim \mathcal{N}(0,50)$.

We study the AR($4$) SV model  to demonstrate the efficiency gain achieved by the proposed CPM method operating in the univariate disturbance space over the CPM method operating in the four-dimensional conventional SS.
The simulation was carried out for two cases with $T=1000$ and $T=3000$. The CPM methods (with particle sorting) was
 omitted in the $T=3000$ case because of the computational complexity involved in sorting the multidimensional particles.

Table~\ref{tab:sv1000} reports the estimation results.
The corresponding trace plots and the ACFs are shown in Figures~\ref{fig:sim:sv4:1000} and~\ref{fig:sim:sv4:3000}.
The traceplots show that the six PM samplers converge adequately to the same invariant distribution.

We note that the estimated correlation of the logs of the likelihood in the CPM method decreases with the dimension of the
state:  a random number correlation $\rho=0.9999$ induced only $\rho_l \approx 0.7$ correlation within the log likelihood estimates in the AR($4$) SV model, while for a univariate SV model we obtain $\rho_l > 0.99$.

Our results show that the proposed CPM operating in the disturbance space is able to target almost the same log likelihood variance as targeted by the standard CPM method operating on the original state space. That is, the proposed CPM method uses as many particles as required by the standard  CPM method and as few as 15\% of that required by the standard PM method.
This validates the key claim of this paper that sorting the disturbances is sufficient
 to maintain order within the particles in the way that it preserves the log likelihood correlation as much as that preserved by directly sorting the multidimensional particles. Moreover the MCMC convergence obtained by CPM method using the proposed multidimensional Euclidean sorting is commensurate with that of the Hilbert curve method. Hence the proposed Euclidean sorting scheme is valid and applicable to the CPM scheme. It can also be observed from the table that the proposed CPM method for the disturbance SS models is $\approx 2$ times and $\approx 9$ times better than the standard PM in terms of RTNV for $T=1000$ and $T=3000$ respectively. A tremendous gain over the conventional CPM method using the Hilbert curve and Euclidean sorting can also be observed: $\approx 12$ times better than CPM with Euclidean sorting and $\approx 30$ better than CPM with Hilbert sorting at $T=1000$. This improvement will increase with the target state dimensionality and the number of observations.
It is noteworthy that although the conventional CPM method operating in the target space facilitates the use of fewer particles for parameter estimation, the method does not sustain its gain in terms of RTNV, at small values of T, due to the inclusion of the resource hungry multidimensional sorting procedure. This advocates the need for designing resource efficient multidimensional sorting methods or improved filtering schemes, e.g., the efficient importance sampling \cite{JeanEIS2007,Scharth2016}, that do not require frequent resampling of the particles.

The best method in this study is clearly the BPM method. It can be observed that the BPM method with ARW is $\approx 11$ times and $\approx 21$ times better than the proposed CPM method for the disturbance SS models and the standard PM method respectively at $T=1000$, and $16$ times and a stupendous $140$ times better at $T=3000$. The BPM with gradient information in the proposal exhibits improvement in the IACT at $T=3000$.

\begin{table}[H]
\centering
$
\begin{array}{c|cccccc}
\text{PM method}&(a)&(b)&(c)&(d)&(e)&(f)\\
\text{Proposal}&\text{ARW} &\text{ARW} &\text{ARW} &\text{ARW} &\text{ARW} &\text{Gradient}\\
\hline \hline
{\bf T=1000} & &  & & & &\\
N & 500 & 75  & 75 & 75 & 50 & 50 \\
\text{Var}(\log\wh L(\boldsymbol\theta)) & \approx1 & 5.4515& 5.4515& 5.4515 &0.5484 & 0.5484\\
\rho & & 0.9999 & 0.9999 & 0.9999 & & \\
\rho_l & & 0.7793 &0.7050 &0.7574 & 0.9167 & 0.9167 \\
2.16^2/(1-\rho_l^2) & &11.8827  &9.2760 & 10.9452& \approx 30 & \approx 30 \\ 
\text{IACT}(\phi)& 14.5751& 16.2132 & 18.2750 &11.7675 &9.4895 &16.5919 \\
\text{IACT}(\tau^2)& 11.3062& 15.8541 &19.3443 & 13.8811  &9.7189 &8.6377 \\
\text{E}(\phi)& 0.9852& 0.9809 &0.9862  &0.9852 &0.9843 &0.9863 \\
\text{E}(\tau^2)& 0.1087&0.1228  &0.1138  &0.1175 &0.1143 &0.1223 \\
\text{STD}(\phi)& 0.0077& 0.0085 &0.0073  & 0.0078 &0.0074 & 0.0062\\
\text{STD}(\tau^2)& 0.0346&0.0405  &0.0355 & 0.0366 &0.0342 & 0.0291\\
\text{Acc. Rate} &0.2219 & 0.2286 & 0.2253 &0.2274 & 0.2321&0.5465 \\ 
\overline{\text{IACT}}& 12.9407& 16.0337 & 18.8096 &12.8243& 9.6042& 12.6148\\
\text{Time} (s) & 0.3440 & 3.5802 &  1.3894  & 0.1723  & 0.0220  & 0.0261 \\
\text{TNV} &4.4516 &  57.4038 &  26.1341  &  2.2096   & 0.2113   & 0.3292\\
\text{RTNV} & 1.0000 &  12.8951  &  5.8707  &  0.4964   & 0.0475 &   0.0740 \\ \hline \hline
{\bf T=3000} & &  & & & &\\
N & 1000 &  & & 100 & 30 & 50\\
\text{Var}(\log\wh L(\boldsymbol\theta)) & \approx1 &  & & 8.0603 & 2.6710 & 1.7348\\
\rho & &  & & 0.9999 & & \\
\rho_l & &  & & 0.7672 & 0.9167 & 0.9167 \\
2.16^2/(1-\rho_l^2) & &  & & 11.3414 & \approx 30 & \approx 30 \\ 
\text{IACT}(\phi)&11.76 &  &  & 15.26 & 19.4 & 12.1\\
\text{IACT}(\tau^2)&16.05 &  &  & 11.47 & 16.96 & 12.74\\
\text{E}(\phi)&0.974 &  &  & 0.9742 & 0.9736 & 0.9742\\
\text{E}(\tau^2)&0.1067 &  &  & 0.1083 & 0.1131 & 0.1087\\
\text{STD}(\phi)&0.0077 &  &  & 0.0075 & 0.0070 & 0.0063\\
\text{STD}(\tau^2)&0.0202 &  &  & 0.0200 & 0.0192 & 0.0165\\
\text{SE}(\phi)&0.0002 &  &  & 0.0002 & 0.0003 & 0.0002\\
\text{SE}(\tau^2)&0.0007 &  &  & 0.0005 & 0.0006 & 0.0005\\
\text{Acc. Rate} &0.2295 &  &  & 0.2264 & 0.2239 & 0.3643 \\ 
\overline{\text{IACT}}&13.9081 &  &  & 13.3636 & 18.1790 & 12.4221\\
\text{Time} (s) & 5.6734 &  &  & 0.6564 & 0.0312 & 0.0351 \\
\text{TNV} & 78.9060  &   &&  8.7719  &  0.5672  &  0.4360 \\
\text{RTNV} & 1.0000 &  &  & 0.1112 &  0.0072 & 0.0055
\end{array}
$
\vspace*{0.15in}
\caption{Estimation results for the AR($4$) SV model. The columns correspond to, (a) the standard PM, (b) the CPM based on the conventional PF with sorting the multidimensional particles using the Hilbert curves method, (c) the CPM based on the conventional PF with sorting the multidimensional particles using the proposed Euclidean sorting method, (d) the proposed CPM based on the disturbance PF with sorting the univariate disturbances, (e) the proposed BPM based on the disturbance PF, and (f) the proposed BPM based on the disturbance PF with the proposal using the derivative information. Time is the elapsed time required to complete one MCMC iteration.}
\label{tab:sv1000}
\end{table}

\subsection{Nonstationary growth model}
The univariate nonlinear nonstationary growth model is described by the SS model
\begin{gather}
x_{t} =\frac{x_{t-1}}{2}+25\frac{x_{t-1}}{1+x_{t-1}^{2}}+8\cos\left(1.2t\right)+a_{t} \\
y_{t} =\frac{x_{t}^{2}}{20}+e_{t},
\end{gather}
where $x_{1}\sim N\left(0,5\right)$, $a_{t}\sim N\left(0,\tau^{2}\right)$, and $e_{t}\sim N\left(0,\sigma^{2}\right)$. The parameters for this model are $\boldsymbol{\theta}=\left(\tau^{2},\sigma^{2}\right)$. This is a popular example used in the literature to assess the performance of the particle filter for nonlinear models.
The number of observations is set to $T=200$. The true parameter values are $\tau^2_{\text{true}}=10$ and $\sigma^2_{\text{true}}=1$. The priors for the parameters  $\text{log } \tau^2 \sim \mathcal{N}(0,50)$ and $\text{log } \sigma^2 \sim \mathcal{N}(0,50)$.
Table~\ref{tab:growth} summarizes the estimation results. Figure~\ref{fig:sim:growth} shows the corresponding trace plots and the ACFs  We
observe that the four PM samplers converge adequately to the same invariant distribution.

Note that in the CPM method we were unable to target the optimal variance.
Our tests revealed that, with the number of particles $N$ chosen such that $\text{Var}(\log\wh L(\boldsymbol\theta)) \approx 2.16^{2}/\left(1-\rho_{l}^{2}\right)$ with $\rho_l=0.9868$, the MCMC chain is sticky and the IACT value is large. This behavior is observed consistently in nonstationary models with multivariate states, e.g., the spline model, the bearings only tracking model, etc. Hence in our study, we increased the number of particles so that the MCMC chain mixes well.
In contrast, our tests  revealed that the proposed BPM method applies to a general class of SS models and induces a precisely known and controllable (via the number of parallel cores) correlation within the log likelihood estimates.

The table shows that in terms of the TNV, the BPM method is $\approx 2.5$ times better than CPM and the standard PM methods. Note that including gradient information in the proposal improves the efficiency of the BPM sampler in terms of the IACT and the method is $\approx 1.5$ times better than the ARW BPM method.  It is possible that increasing the number of particles will make the gradient more informative (i.e., reduce its variance) and will hence improve the parameter proposal.

\begin{table}
\centering
$
\begin{array}{c|cccc}
\text{PM method}&\text{Standard PM} &\text{CPM} & \text{BPM} &\text{BPM}\\
\text{Proposal}&\text{ARW} &\text{ARW} &\text{ARW} &\text{Gradient}\\
\hline \hline
N & 2000 & 1000 & 400 & 400 \\
\text{Var}(\log\wh L(\boldsymbol\theta)) & \approx1 & 1.9007 & 10.7900 & 10.7900 \\
\rho & & 0.9999 & & \\
\rho_l & & 0.9868 & 0.9167 & 0.9167 \\
2.16^2/(1-\rho_l^2) & & 177.9890 & \approx 30 & \approx 30 \\
\hline \hline
\text{IACT}(\tau^2)&12.22 & 17.31 &18.3536 & 12.6484\\
\text{IACT}(\sigma^2)&11.38 & 14.01 &13.0572 & 8.6455\\
\text{E}(\tau^2)&11.32 & 11.21 &11.3488 & 11.2791\\
\text{E}(\sigma^2)&0.9375 & 0.9465 &0.9535 & 0.9657\\
\text{STD}(\tau^2)&1.495 & 1.521 &1.5035 & 0.9629\\
\text{STD}(\sigma^2)&0.1974 & 0.1987 &0.2015 &0.1297 \\
\text{SE}(\tau^2)&0.0427 & 0.0517 &0.0526 &0.0280 \\
\text{SE}(\sigma^2)&0.0054 & 0.0061 & 0.0060& 0.0031\\
\text{Acc. Rate} &0.231 & 0.2166 &0.2335 &0.4364 \\
\hline \hline
\overline{\text{IACT}}&11.8000 &  15.6585  &15.7054 &10.6470 \\
\text{Time Time} (s) &0.0868&  0.0571&0.0255  & 0.0273\\
\text{TNV} & 1.0242 & 0.8941 &0.4004 & 0.2906\\
\text{RTNV} & 1.0000 & 0.8729 &0.3910 &0.2837
\end{array}
$
\vspace*{0.15in}
\caption{Estimation results for the PM methods for $T=200$ for the nonstationary growth model. The columns correspond to, (a) the standard PM, (b) the CPM with univariate particles sorted, (c) the proposed BPM method, and (d) the proposed BPM method with the PM proposal using the derivative information. Time is the elapsed time required to complete one MCMC iteration.}
\label{tab:growth}
\end{table}
\subsection{Cubic spline model}
\label{model:sim:spline}
A simple bivariate cubic spline model is defined by the SS formulation
\begin{align}
\mathbf{x}_{t}=F (\delta_{t-1}) \;  \mathbf{x}_{t-1} + a_{t}, \\
y_{t} = \left[\begin{array}{cc}
1 & 0\end{array}\right] \mathbf{x}_{t} + e_{t}
\end{align}
where the state transition matrix is
$F\left(\delta_{t-1}\right)=\left[\begin{array}{cc}
1 & \delta_{t-1}\\
0 & 1
\end{array}\right]$, the process noise is $a_{t} \sim \mathcal{N}(0,\tau^{2} \; U(\delta_{t-1}))$ with
$U\left(\delta_{t-1}\right) = \left[\begin{array}{cc}
\delta_{t-1}^{3}/3 & \delta_{t-1}^{2}/2\\
\delta_{t-1}^{2}/2 & \delta_{t-1}
\end{array}\right]$ and the noise variance is $e_{t} \sim \mathcal{N}(0,\sigma^2)$. The rate at which observations are received is chosen to be $\delta_{t}=1/T$.
The model parameters  are $\boldsymbol{\theta}=\left(\tau^{2},\sigma^{2}\right)$.

The number of observations is set to $T=500$. The true parameter values are $\tau^2_{\text{true}}=4$ and $\sigma^2_{\text{true}}=0.25$. The priors for the parameters are $\tau^{2}\sim \text{IG}(1,1)$ and $\sigma^{2}\sim \text{IG}(1,1)$. The BPM method with gradient information is not included in this study because the gradient involves the inversion of the matrix $U\left(\delta_{t-1}\right)$, which has a large value, thus making the gradient computation numerically unstable.

Table~\ref{tab:spline} summarizes the estimation results. Figure~\ref{fig:sim:spline} shows the corresponding trace plots and the ACFs.
The figure shows that the PM samplers converge adequately to the same invariant distribution and that the estimated mean and standard deviation values are similar. The results also show that
 the BPM method also exhibits improved performance in terms of IACT and is $\approx 7$ times better than the standard PM method.

\begin{table}
\centering
$
\begin{array}{c|cc}
\text{PM method}&\text{Standard PM} &\text{BPM} \\
\text{Proposal}&\text{ARW} &\text{ARW} \\
\hline \hline
\# N & 2000 & 80 \\
\text{Var}(\log\wh L(\boldsymbol\theta)) & \approx1 & 1.6439  \\
\rho_l & & 0.9167 \\
2.16^2/(1-\rho_l^2) & & \approx 30 \\ \hline \hline
\text{IACT}(\tau^2)&9.0207 &28.0079 \\
\text{IACT}(\sigma^2)&8.7280 & 15.3553 \\
\text{E}(\tau^2)&2.3308 & 2.3222 \\
\text{E}(\sigma^2)&0.2341 &0.2343 \\
\text{STD}(\tau^2)&0.8221 & 0.8441 \\
\text{STD}(\sigma^2)&0.0156 & 0.0161 \\
\text{SE}(\tau^2)&0.0116 & 0.0364 \\
\text{SE}(\sigma^2)&2.17 \cdot 10^{-4} & 5.2\cdot 10^{-4} \\
\text{Acc. Rate}&0.2340 & 0.2262 \\ \hline \hline
\overline{\text{IACT}}& 8.8743 & 21.6816 \\
\text{Time} (s) & 0.3854 & 0.0244 \\
\text{TNV} & 3.4202  &  0.5290 \\
\text{RTNV} &1.0000  & 0.1546
\end{array}
$
\vspace*{0.15in}
\caption{Estimation result for the PM methods for $T=500$ for the bivariate cubic spline model. The columns correspond to, (a) the standard PM, and (b) the BPM method. Time is the elapsed time required to complete one MCMC iteration}
\label{tab:spline}
\end{table}

\subsection{Lotka Volterra model}
The Lotka Volterra model is a reaction network that models the interaction between species \citep{Golightly2011}. Here we study a simple bivariate nonstationary stochastic Lotka Volterra model. Define the state vector by $\mathbf{x}=(x_{1}, x_{2})^{\top}$, where $x_{1}$ denotes the prey and $x_{2}$ the predators. The predator prey system is comprised of three reactions
\begin{align*}
\mathcal{R}_{1} &: \mathcal{X}_{1}  \longrightarrow 2 \; \mathcal{X}_{1} ,\quad
\mathcal{R}_{2} : \mathcal{X}_{1} + \mathcal{X}_{2} \longrightarrow 2 \; \mathcal{X}_{2} , \quad
\mathcal{R}_{3} : \mathcal{X}_{2}  \longrightarrow \emptyset
\end{align*}
representing the prey reproduction, the predator-prey interaction and the predator death. The rate constant vector for the three reactions is defined as ${\bf c}=(c_{1},c_{2},c_{3})^{\top}$ and the associated hazard function is
$
h(\mathbf{x},{\bf c})=(c_{1} x_{1,t}, \; c_{2} x_{1,t} x_{2,t}, \; c_{3} x_{2,t})^{\top}.
$

The stoichiometry matrix used to update the state if a reaction occurs is given by
\[
S=\left(\begin{array}{ccc}
1 & -1 & 0\\
0 & 1 & -1
\end{array}\right)
\]
We set $\mathbf{x}_{1}=(100,100)^{\top}$ and the (intractable) forward simulations are conducted using the Gillespie algorithm \cite{Wilkinson2011book}. The observation equation is
\begin{align}
\mathbf{y}_{t} = \mathbf{x}_{t} + e_{t},
\end{align}
where the noise variance $e_{t}\sim \mathcal{N}(0,\sigma^{2})$.
The parameters for this model are
$\boldsymbol\theta=\{{\bf c},\sigma^2\}$.

The number of observations is $T=50$. The true parameter values are ${\bf c}_{\text{true}}=(0.5, 0.0025, 0.3)$ and $\sigma^2_{\text{true}}=0.5$. The priors for the parameters are $c_{i=1:3} \sim \mathcal{U}(0,1)$ and $\sigma^{2}\sim \text{IG}(1,1)$.
Table~\ref{tab:predatorprey} summarizes the estimation results. Figure~\ref{fig:sim:lv} plots
the corresponding trace plots and the ACFs, and shows  that the two PM samplers converge adequately to the same invariant distribution.
 The results suggest that the  BPM method exhibits improved performance in terms of IACT and is $\approx 40$ times better than the standard PM method.

\begin{table}
\centering
$
\begin{array}{c|cc}
\text{PM method}&\text{Standard PM} &\text{BPM} \\
\text{Proposal}&\text{ARW} &\text{ARW} \\
\hline \hline
N & 600 & 30 \\
\text{Var}(\log\wh L(\boldsymbol\theta)) & \approx1 & 1.8400 \\
\rho_l & & 0.9167 \\
2.16^2/(1-\rho_l^2) & & \approx 30 \\ \hline \hline
\text{IACT}(c_{1})&20.65 & 26.1394\\
\text{IACT}(c_{2})&21.68 &21.0507  \\
\text{IACT}(c_{3})&19.7 &28.7471  \\
\text{IACT}(\sigma^2)&22.97 &22.0728 \\
\text{E}(c_{1})&0.5728 &0.5999  \\
\text{E}(c_{2})&0.0017 & 0.0016\\
\text{E}(c_{3})&0.2917 & 0.3130\\
\text{E}(\sigma^2)&0.7622 & 0.9211 \\
\text{STD}(c_{1})&0.1274 &0.1200 \\
\text{STD}(c_{2})&0.0011 & 0.0010\\
\text{STD}(c_{3})&0.1203 & 0.1065 \\
\text{STD}(\sigma^2)&0.1678 &0.1831 \\
\text{SE}(c_{1})&0.0047 &0.0050 \\
\text{SE}(c_{2})&4.18\cdot 10^{-5} &3.79 \cdot 10^{-5} \\
\text{SE}(c_{3})&0.004 & 0.0047\\
\text{SE}(\sigma^2)&0.0066 &0.0070\\
\text{Acc. Rate}&0.2263 &0.2271 \\  \hline \hline
\overline{\text{IACT}}&21.2505& 24.5025\\
\text{Time} (s) &6.4032 & 0.1462 \\
\text{TNV} &136.0710 & 3.5823  \\
\text{RTNV} &1.0000 & 0.0263
\end{array}
$
\vspace*{0.15in}
\caption{Estimation result of the PM methods for $T=50$ for the Lotka Volterra model. The columns correspond to, (a) the standard PM, and (b) the  BPM method operating on $12$ parallel cores. Time is the elapsed time required to complete one MCMC iteration.}
\label{tab:predatorprey}
\end{table}

\section{Application of the stochastic volatility model to financial returns data}\label{sec:real_data}
This section studies the performance of the proposed PM methods in financial data. We consider two data sets corresponding to, (a) the daily returns for the Australian All Ordinaries Price Index (All Ords) between 26/11/2002 and 04/08/2016, and (b) the daily returns for the American stock market index S\&P $500$ index between 05/01/1970 and 22/11/2016. The prices are converted to daily logarithmic returns as follows
\begin{align*}
r_{t}=\log\left(\frac{\text{price}_{t}}{\text{price}_{t-1}}\right)\times100 .
\end{align*}
The first data set is the All Ords index and contains $T=3564$ observations. The
second data set corresponds to the S\&P$500$ index and contains $T=7832$ observations. The data is retrieved from Yahoo finance.
We fit the AR($4$) SV model described in section~\ref{model:sim:sv4} to these two data sets. The parameters for the models are $\boldsymbol{\theta}=\left(\phi,\tau^{2}\right)$. The priors for the parameters and the ARW proposal are the same as described in section~\ref{model:sim:sv4}.

Table~\ref{tab:sv:finance3564} reports the estimation results corresponding to the All Ords data set,
for the standard PM, disturbance CPM and the BPM methods.
 The BPM method was implemented on $G=12$ independent cores.
The result corresponding to the CPM with particle sorting is omitted due to the computational infeasibility of the method.
Figure~\ref{fig:sim:sv4:3564} shows the corresponding trace plots and the ACFs, and shows that
the PM samplers converge adequately to the same invariant distribution.
The disturbance CPM method uses $4.3$ times fewer particles and is $26$ times superior to the standard PM method. The clear winner is the BPM method. The BPM (with an ARW proposal) method uses $26$ times fewer particles than the standard PM method, and is $\approx 838$ times and $\approx 33$ times superior to the standard PM and the disturbance CPM methods respectively. Note that the number of particles used in the BPM with the proposal using the derivative information is slightly increased to make the gradient more informative. This leads to reduced $\overline{\text{IACT}}$ values. It can be observed that the BPM method (with the proposal using the Hessian information) is $\approx 1540$ times and $\approx 57$ times superior to the standard PM and the disturbance CPM methods respectively.

\begin{table}
\centering
$
\begin{array}{c|cccc}
\text{PM method}&\text{Standard PM}&\text{Disturb CPM}&\text{BPM}&\text{BPM}\\
\text{Proposal}&\text{ARW}&\text{ARW}&\text{ARW}&\text{Gradient}\\
\hline \hline
\text{{\bf All Ords, T=3564}}&&&&\\
N & 1600 & 230  & 80 & 100 \\
\text{Var}(\log\wh L(\boldsymbol\theta)) & \approx1 &10.6833 & 1.6335&1.6215 \\
\rho & & 0.9999 & & \\
\rho_l & &0.7723 &0.9167 & 0.9167 \\
2.16^2/(1-\rho_l^2) & &11.5652  & \approx 30 & \approx 30 \\ 
\text{IACT}(\phi)&13.92 & 11.9 & 13.2 & 7.73\\
\text{IACT}(\tau^2)&12.53 & 12.61 & 13.94 & 8.235\\
\text{E}(\phi)&0.9831 & 0.9834 & 0.9825 & 0.9831\\
\text{E}(\tau^2)&0.0848 & 0.0836 & 0.092 & 0.0914\\
\text{STD}(\phi)&0.0056 & 0.0053 & 0.0054 & 0.0046\\
\text{STD}(\tau^2)&0.0164 & 0.0168 & 0.0173 & 0.0139\\
\text{SE}(\phi)&0.0002 & 0.0002 & 0.0002 & 0.0001\\
\text{SE}(\tau^2)&0.0005 & 0.0005 & 0.0005 & 0.0003\\
\text{Acc. Rate} &0.2263 & 0.2313 & 0.2274 & 0.3759 \\ 
\overline{\text{IACT}} &13.2259 &12.2581 &13.5682 &7.9821  \\
\text{Time Time} (s) & 39.2135 & 1.5705 & 0.0456 & 0.0422  \\
\text{TNV} & 518.6358& 19.2514& 0.6187& 0.3368 \\
\text{RTNV} &1.0000 &0.0371 &0.0012 &0.0006  \\ \hline \hline
\text{{\bf S\&P 500, T=7832}}&&&&\\
N & 3000 & 500  & 250 & 250 \\
\text{Var}(\log\wh L(\boldsymbol\theta)) & \approx1 &  8.7698 &2.2851 & 2.2851 \\
\rho & & 0.9999 & & \\
\rho_l & & 0.7471 &0.7500 & 0.7500 \\
2.16^2/(1-\rho_l^2) & & 10.5587 & 10.6642 & 10.6642 \\ 
\text{IACT}(\phi)&&13.66 & 16.3 &5.1451 \\
\text{IACT}(\tau^2)&&13.2 & 15.58 &6.3945 \\
\text{E}(\phi)&&0.9752 & 0.9753 &0.9750 \\
\text{E}(\tau^2)&&0.0901 & 0.0914 &0.0926 \\
\text{STD}(\phi)&&0.0050 & 0.0053 &0.0040 \\
\text{STD}(\tau^2)&&0.0126 & 0.0129 &0.0105 \\
\text{SE}(\phi)&&0.0002 & 0.0002 & 8.98 \cdot 10^{-5} \\
\text{SE}(\tau^2)&&0.0004 & 0.0005 & 0.0003 \\
\text{Acc. Rate}&&0.2262 & 0.2317 & 0.4428 \\ 
\overline{\text{IACT}} & &13.4335 &15.9397 & 5.7698 \\
\text{Time Time} (s) & &17.9041 &0.1578 &0.1517 \\
\text{TNV} & &240.5150 &2.5153 &0.8752 \\
\text{RTNV} & & 1.0000 &0.0105 &0.0036
\end{array}
$
\vspace*{0.15in}
\caption{Estimation result of the PM methods for the financial returns data applied to the AR($4$) SV model. Time is the elapsed time required to complete one MCMC iteration.}
\label{tab:sv:finance3564}
\end{table}

Table~\ref{tab:sv:finance3564} summarizes the estimation results corresponding to the S\&P$500$ data with $T=7832$,
for the disturbance based CPM and the BPM methods.
The standard PM requires $3000$ particles to satisfy $\text{Var}(\log\wh L(\boldsymbol\theta)) \approx 1$. Consequently its TNV is large. The results corresponding to the standard PM and the CPM with particle sorting are excluded due to the computational infeasibility of the methods.
Figure~\ref{fig:sim:sv4:7832} reports the corresponding trace plots and the ACFs  and shows
that the PM samplers converge adequately to the same invariant distribution.
The BPM method is implemented on $G=4$ independent cores.  The BPM methods are $\approx 100$ and $\approx 280$ times superior to the disturbance CPM method using only half the number of particles used in the CPM. The BPM method using the gradient information are $\approx 3$ times better than the ARW BPM method.

\section{Conclusion}
\label{sec:concl}
The contribution of this paper is two-fold, (a) a CPM method for parameter estimation in disturbance SS models is proposed, and (b) a BPM method for general time series models is proposed. The main insight in the CPM method for disturbance SS models is that the disturbance SS possesses the same properties as that the conventional SS as required to preserve the correlation within the likelihood estimates. This insight is the basis for the proposed CPM method for disturbance SS models.
The key innovation of this proposal is that the CPM is performed in the disturbance space rather than the SS by virtue of preserving the likelihood correlation by maintaining similarity in the disturbance states as opposed to the traditional approach that maintains similarity in the target states. The main insight in the BPM method is that implementing multiple independent particle filters, with the likelihood estimated by averaging the likelihood estimates obtained from
separate particle filters, facilitates the reduction in the variance of the log of the estimated
likelihood  and hence in the use of fewer particles. This insight is the basis for the design of a flexible means to block the random numbers in order to obtain a desired correlation for the log of likelihood estimates.
\bibliography{refs01}   
\bibliographystyle{apalike}

\begin{appendices}\label{appndx}

\section{Proposal using information from the derivatives}
\label{appdx:derivatives}
This section presents the construction of the proposal density in MCMC that makes use of the derivatives of the log likelihood. \cite{Poyiadjis2011} were the first to show how the particle filter methods can be used to estimate the derivatives of the log likelihood for state space models. Their methods might suffer from a computational cost that is quadratic in the number of particles, \cite{Nemeth2015} proposed an alternative method whose computational cost is linear in the number of particles.
They use a combination of kernel density estimation and Rao-Blackwellisation to reduce the Monte Carlo error of the estimates.
For non-linear and non-Gaussian state space models it is impossible to obtain the exact derivatives,
but they can be approximated using the particle approximation of $p\left(\mathbf{x}_{1:T}|\mathbf{y}_{1:T},\boldsymbol{\theta}\right)$.

By Fisher's identity \citep{Cappe2005}
\begin{align}
\nabla\log L\left(\boldsymbol{\theta}\right)=\nabla\log p\left(\mathbf{y}_{1:T}|\boldsymbol{\theta}\right)=\int\nabla\log p\left(\mathbf{x}_{1:T},\mathbf{y}_{1:T}|\boldsymbol{\theta}\right)p\left(\mathbf{x}_{1:T}|\mathbf{y}_{1:T},\boldsymbol{\theta}\right) d \mathbf{x}_{1:T}.
\end{align}
where
\begin{align}
\nabla\log p\left(\mathbf{x}_{1:T},\mathbf{y}_{1:T}|\boldsymbol{\theta}\right)=\sum_{t=1}^{T}\left\{ \nabla\log f\left(\mathbf{y}_{t}|\mathbf{x}_{t},\boldsymbol{\theta}\right)+\nabla\log g\left(\mathbf{x}_{t}|\mathbf{x}_{t-1},\boldsymbol{\theta}\right)\right\}.
\end{align}
Similarly, the observed negative Hessian (second derivative) matrix satisfies Louis' identity \citep{Louis1982}
\begin{align}
-\nabla^{2}\log L\left(\boldsymbol{\theta}\right)=\nabla\log L\left(\boldsymbol{\theta}\right)\nabla\log L\left(\boldsymbol{\theta}\right)^T-\frac{\nabla^{2}L\left(\boldsymbol{\theta}\right)}{L\left(\boldsymbol{\theta}\right)}\label{eq:hessian1}
\end{align}
where
\begin{eqnarray}
\frac{\nabla^{2}L\left(\boldsymbol{\theta}\right)}{L\left(\boldsymbol{\theta}\right)} & = & \int\nabla\log p\left(\mathbf{x}_{1:T},\mathbf{y}_{1:T}|\boldsymbol{\theta}\right)\nabla\log p\left(\mathbf{x}_{1:T},\mathbf{y}_{1:T}|\boldsymbol{\theta}\right)^{T}p\left(\mathbf{x}_{1:T}|\boldsymbol{\theta},\mathbf{y}_{1:T}\right) d\mathbf{x}_{1:T},\label{eq:hessian2}\\
 &  & +\int\nabla^{2}\log p\left(\mathbf{x}_{1:T},\mathbf{y}_{1:T}|\boldsymbol{\theta}\right)p\left(\mathbf{x}_{1:T}|\boldsymbol{\theta},\mathbf{y}_{1:T}\right)d\mathbf{x}_{1:T}\nonumber
\end{eqnarray}
and
\begin{align}
\nabla^{2}\log p\left(\mathbf{x}_{1:T},\mathbf{y}_{1:T}|\boldsymbol{\theta}\right)=\sum_{t=1}^{T}\left\{ \nabla^{2}\log f\left(\mathbf{y}_{t}|\mathbf{x}_{t},\boldsymbol{\theta}\right)+\nabla^{2}\log g\left(\mathbf{x}_{t}|\mathbf{x}_{t-1},\boldsymbol{\theta}\right)\right\}
\end{align}
The procedure of \cite{Nemeth2015} to estimate first and second derivatives is outlined in Algorithm \ref{alg:Algorithm-to-estimate gradient and hessian}.
The coefficient $\lambda$ and $h$ is chosen such that $\lambda^{2}+h^{2}=1$. Setting $\lambda=1$ gives the algorithm in \cite{Poyiadjis2011}.  \cite{Nemeth2015} show that the bias and variance of both score estimate and observed information matrix vary according to $\lambda$. Reducing the value of $\lambda$ increases the bias, but it reduces the Monte Carlo variance of estimates. It is also shown that setting $\lambda \approx 0.95$ will produce an estimate for the score and observed information matrix with linearly increasing variance and minimal bias. Therefore we use $\lambda=0.95$ in all our applications.
The parameter proposal adopted in this paper is similar to the one used in \cite{Dahlin2015Derivatives}
\begin{align}
q\left(\boldsymbol{\theta}^{'}|\boldsymbol{\theta},\mathbf{U}\right)=N\left(\boldsymbol{\theta}+\widehat{G}\left(\boldsymbol{\theta};\mathbf{U}\right),\widehat{H}\left(\boldsymbol{\theta};\mathbf{U}\right)\right),
\end{align}
where $\widehat{G}\left(\boldsymbol{\theta};\mathbf{U}\right)=\frac{1}{2}\widehat{\Sigma}^{-1}\widehat{S}$ and $\widehat{H}\left(\boldsymbol{\theta};\mathbf{U}\right)=\widehat{\Sigma}^{-1}$,
and $\mathbf{U}$ is the set of random numbers used to construct the estimators.

\begin{algorithm}[H]
\caption{Algorithm to estimate Gradient and Hessian Matrix
\label{alg:Algorithm-to-estimate gradient and hessian}}
\begin{itemize}
\item Initialise: set $m_{0}^{\left(i\right)}=0$ and $n_{0}^{\left(i\right)}=0$
for $i=1,...,N$, where $N$ is the number of particles, and $S_{0}=0$
and $B_{0}=0$.
\item At iteration $t=1,...,T$
\begin{itemize}
\item Run the particle filter to obtain $\left\{ \mathbf{x}_{t}^{\left(i\right)}\right\} _{i=1}^{N}$,
$\left\{ k_{i}\right\} _{i=1}^{N}$, and $\left\{ w_{t}^{\left(i\right)}\right\} _{i=1}^{N}$,
where $w_{t}^{\left(i\right)}$ is the weight of particle $i$ at time $t$. $k_{i}$ is the ancestor index of particle $i$ at time $t-1$.
\item Normalise the weights $W_{t}^{\left(i\right)}=\frac{w_{t}^{\left(i\right)}}{\sum w_{t}^{\left(i\right)}}$.
\end{itemize}
\item Update the $m_{t}^{\left(i\right)}$ and $n_{t}^{\left(i\right)}$ as follows
\[
m_{t}^{\left(i\right)}=\lambda m_{t-1}^{\left(k_{i}\right)}+\left(1-\lambda\right)S_{t-1}+\nabla\log g_{\theta}\left(\mathbf{y}_{t}|\mathbf{x}_{t}^{\left(i\right)}\right)+\nabla\log p_{\theta}\left(\mathbf{x}_{t}^{\left(i\right)}|\mathbf{x}_{t-1}^{\left(k_{i}\right)}\right)
\]
and
\end{itemize}
\[
n_{t}^{\left(i\right)}=\lambda n_{t-1}^{\left(k_{i}\right)}+\left(1-\lambda\right)B_{t-1}+\nabla^{2}\log g_{\theta}\left(\mathbf{y}_{t}|\mathbf{x}_{t}^{\left(i\right)}\right)+\nabla^{2}\log p_{\theta}\left(\mathbf{x}_{t}^{\left(i\right)}|\mathbf{x}_{t-1}^{\left(k_{i}\right)}\right)
\]
\begin{itemize}
\item Update the score vector
\[
S_{t}=\sum_{i=1}^{N}W_{t}^{\left(i\right)}m_{t}^{\left(i\right)}
\]
\item The observed negative Hessian matrix can be estimated as
\[
\Sigma_{t}=S_{t}S_{t}^{'}-\sum_{j=1}^{N}W_{t}^{\left(j\right)}\left(m_{t}^{\left(j\right)}m_{t}^{'\left(j\right)}+n_{t}^{\left(j\right)}\right)-h^{2}V_{t}
\]
where $V_{t}=V_{t-1}+\sum_{i=1}^{N}W_{t-1}^{\left(i\right)}\left(m_{t-1}^{\left(i\right)}-S_{t-1}\right)^{'}\left(m_{t-1}^{\left(i\right)}-S_{t-1}\right)$
and $B_{t}=\sum_{i=1}^{N}w_{t}^{\left(i\right)}n_{t}^{\left(i\right)}$ \end{itemize}
\end{algorithm}

\section{Plots of the results presented in sections~\ref{sec:sim_study} and  \ref{sec:real_data}}
This section presents the trace plots and the autocorrelation function (ACF) plots corresponding to the results presented in sections~\ref{sec:sim_study} and \ref{sec:real_data}.
\begin{figure*}[ht]
\hspace*{-0.75in}
\includegraphics[width=1.1\textwidth]{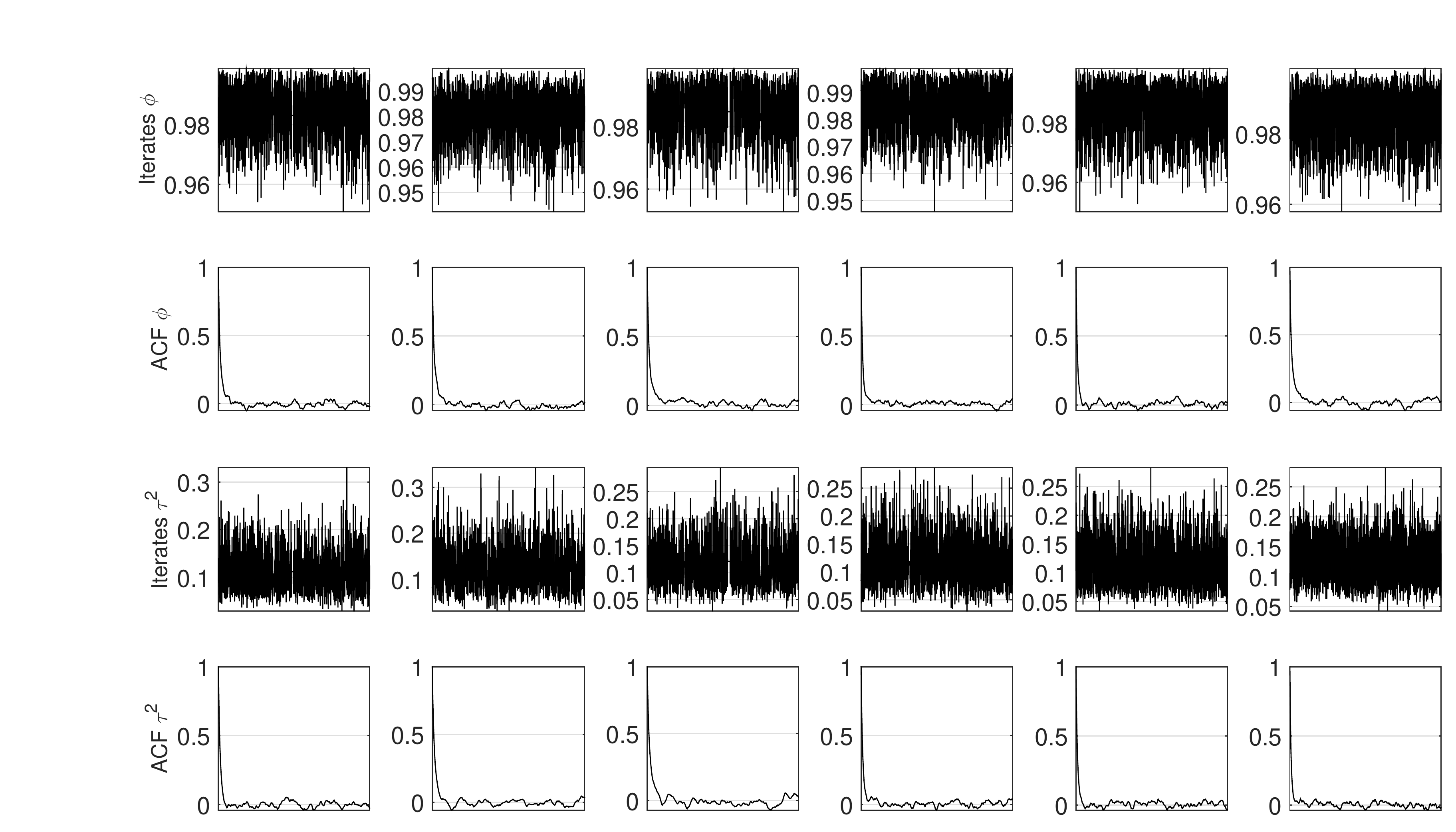}
\vspace*{0.1in}
\caption{The estimation result for the AR($4$) SV model with $T=1000$ for the result reported in Table~\ref{tab:sv1000}. The columns from left to right correspond to, (a) the standard PM, (b) the CPM with sorting the multidimensional particles using the Hilbert curves method, (c) the CPM with sorting the multidimensional particles using the proposed Euclidean sorting method, (d) the CPM with sorting the univariate disturbances, (e) the proposed BPM method, and (f) the BPM method with the proposal using the derivative information.}\label{fig:sim:sv4:1000}
\end{figure*}

\begin{figure*}[ht]
\hspace*{-0.75in}
\includegraphics[width=1.1\textwidth]{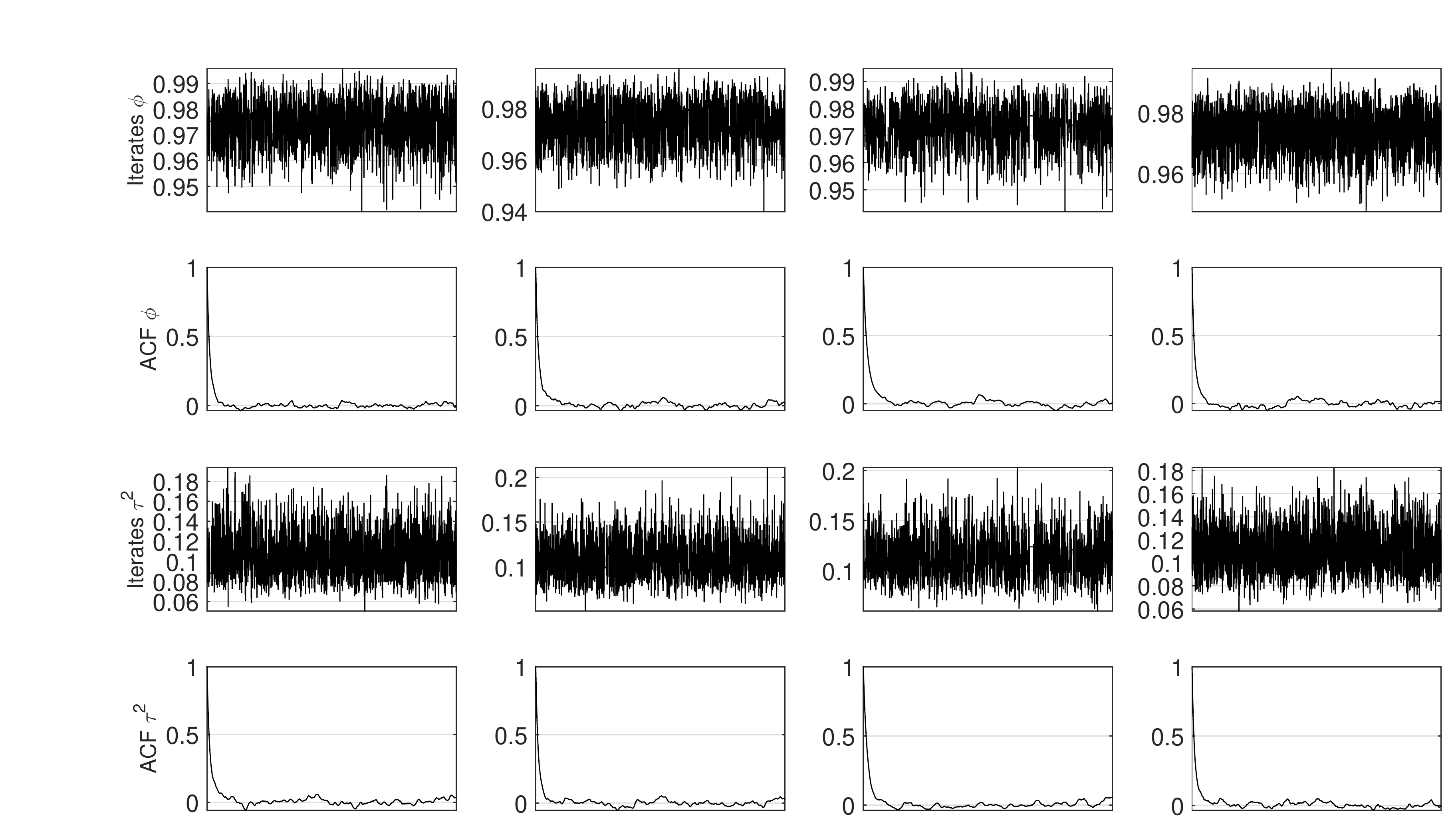}
\vspace*{0.1in}
\caption{The estimation result for the AR($4$) SV model with $T=3000$ for the result reported in Table~\ref{tab:sv1000}. The columns from left to right correspond to, (a) the standard PM, (b) the CPM with sorting the univariate disturbances, (c) the proposed BPM method, and (d) the BPM method with the proposal using the derivative information.}
\label{fig:sim:sv4:3000}
\end{figure*}

\begin{figure*}[ht]
\hspace*{-0.75in}
\includegraphics[width=1.1\textwidth]{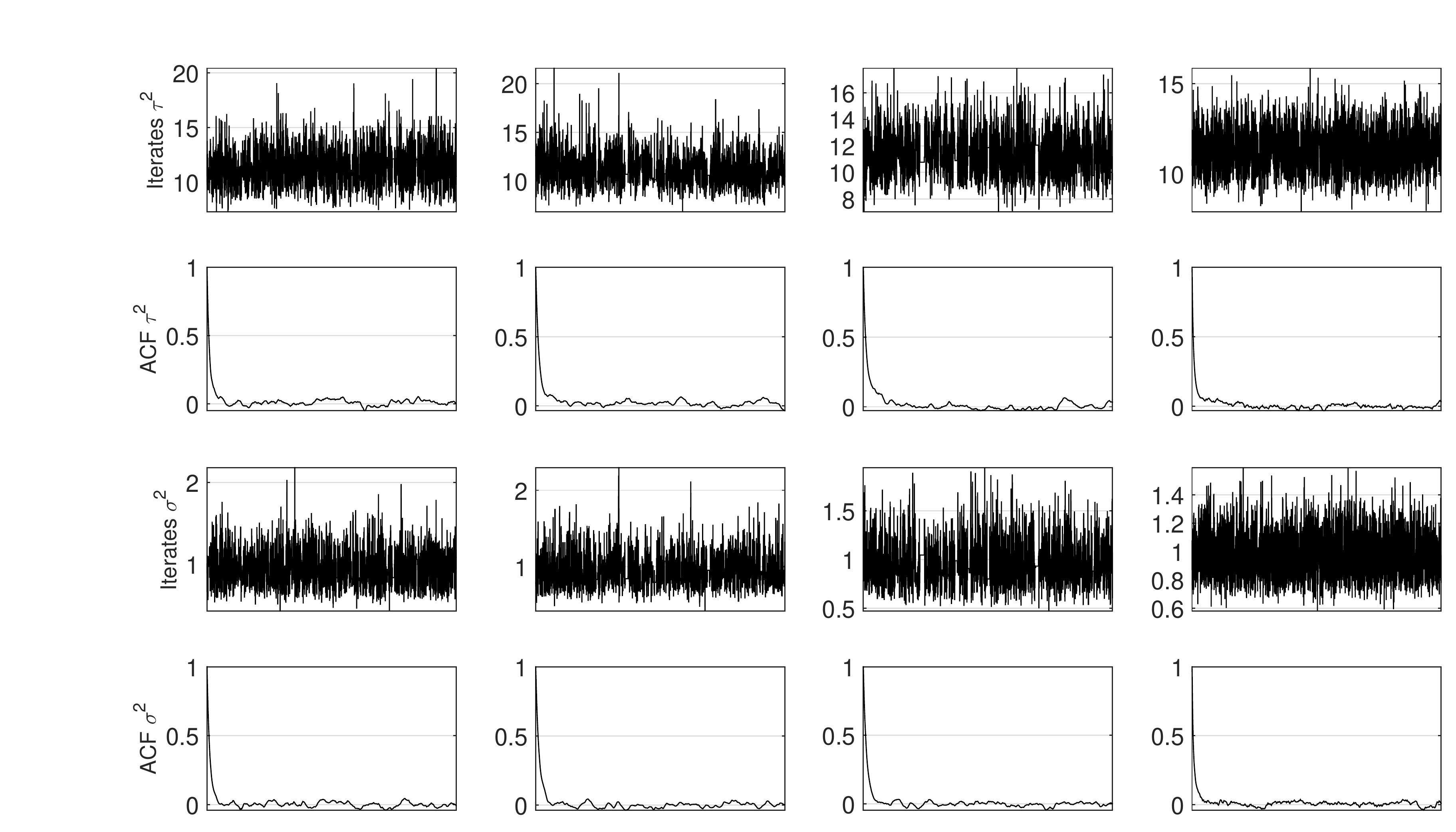}
\vspace*{0.1in}
\caption{The estimation result for the non-stationary growth model corresponding to the result reported in Table~\ref{tab:growth}. The columns from left to right correspond to, (a) the standard PM, (b) the CPM with sorting the univariate particles, (c) the BPM method with RW, and (d) the BPM method with the proposal using the derivative information.}
\label{fig:sim:growth}
\end{figure*}

\begin{figure*}[ht]
\hspace*{-0.75in}
\includegraphics[width=1.1\textwidth]{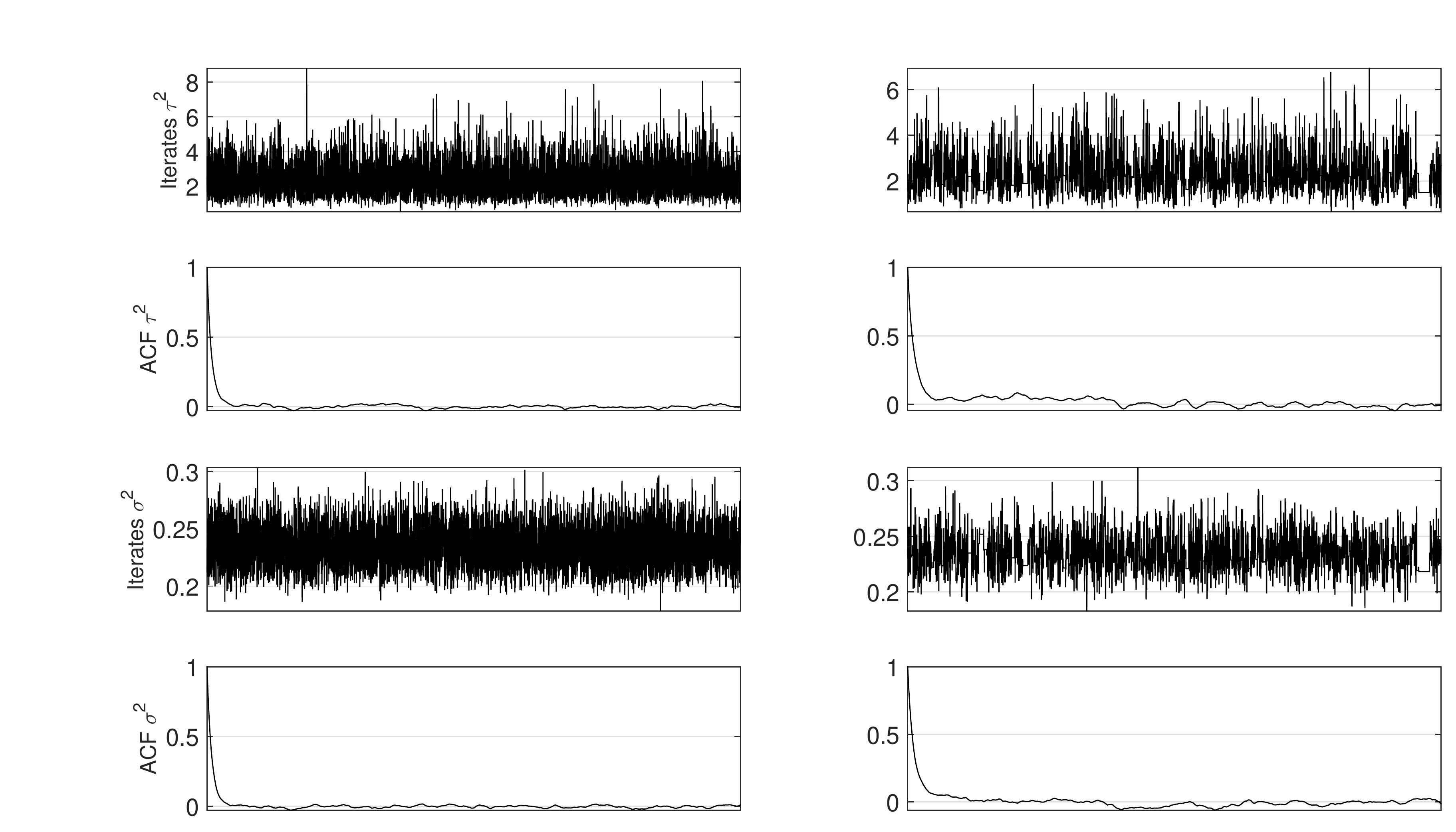}
\vspace*{0.1in}
\caption{The estimation result for the bivariate cubic spline model corresponding to the result reported in Table~\ref{tab:spline}. The columns from left to right correspond to, (a) the standard PM, and (b) the proposed BPM method.}
\label{fig:sim:spline}
\end{figure*}

\begin{figure*}[ht]
\hspace*{-0.75in}
\includegraphics[width=1.1\textwidth]{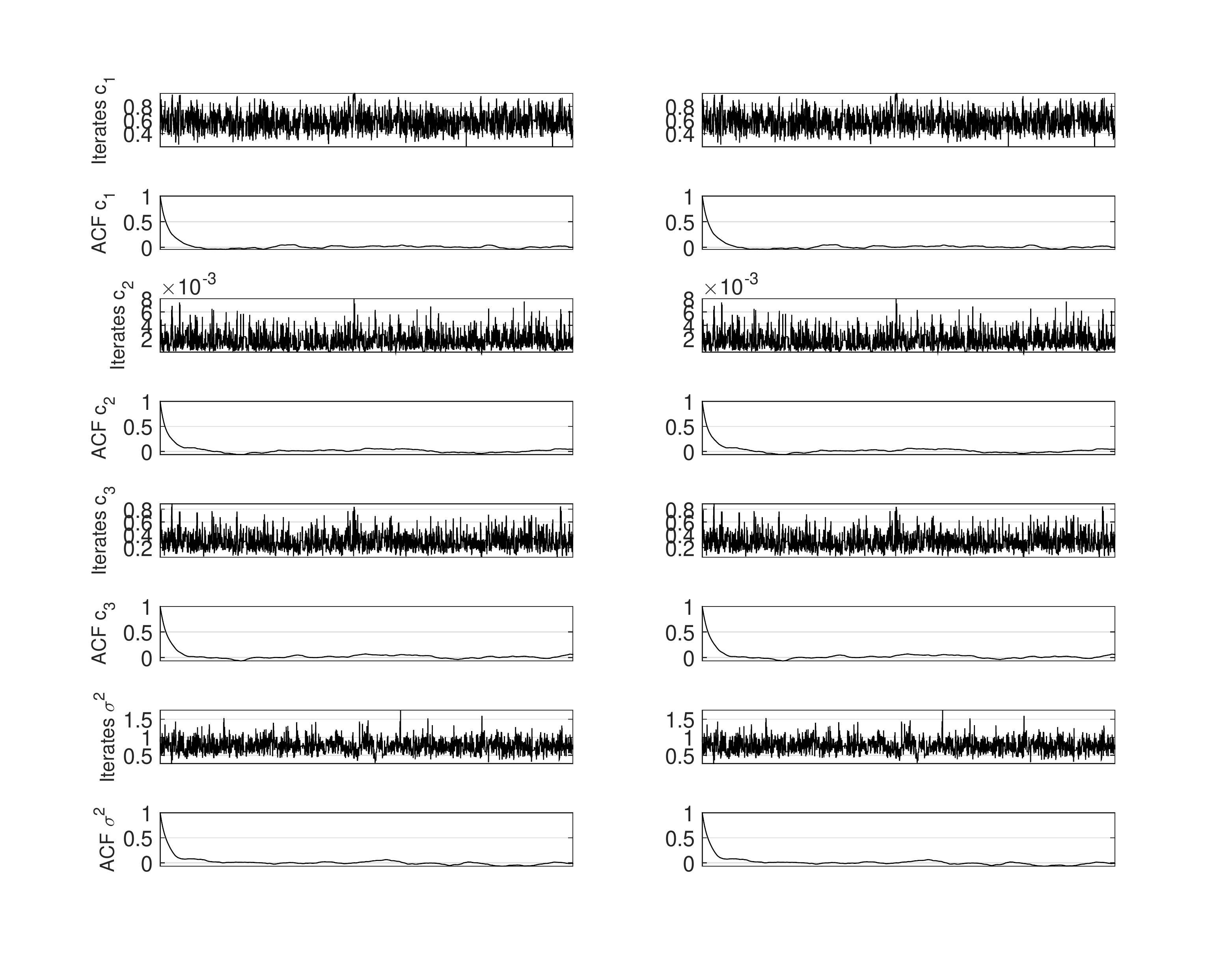}
\vspace*{0.1in}
\caption{The estimation result for the Lotka Volterra model corresponding to the result reported in Table~\ref{tab:predatorprey}. The columns from left to right correspond to, (a) the standard PM, and (b) the BPM method.}
\label{fig:sim:lv}
\end{figure*}

\begin{figure*}[ht]
\hspace*{-0.75in}
\includegraphics[width=1.1\textwidth]{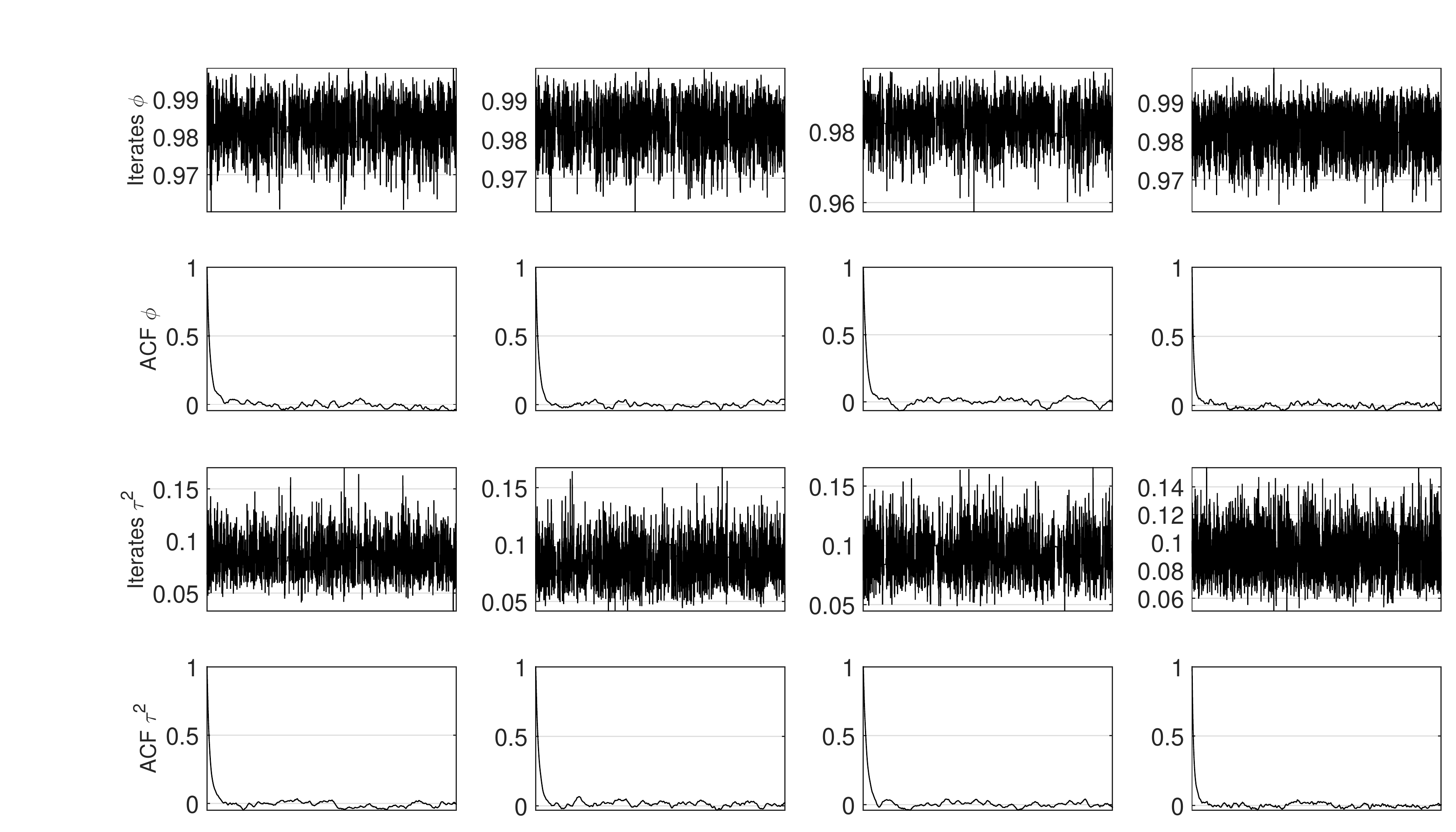}
\vspace*{0.1in}
\caption{The estimation result for the Australian All Ords financial returns data applied to the AR($4$) SV model corresponding to the result reported in Table~\ref{tab:sv:finance3564}. The columns from left to right correspond to, (a) the standard PM, (b) the CPM with sorting the univariate disturbances, (c) the proposed BPM method with RW, and (d) the proposed BPM method with the proposal using the derivative information.}
\label{fig:sim:sv4:3564}
\end{figure*}

\begin{figure*}[ht]
\hspace*{-0.75in}
\includegraphics[width=1.1\textwidth]{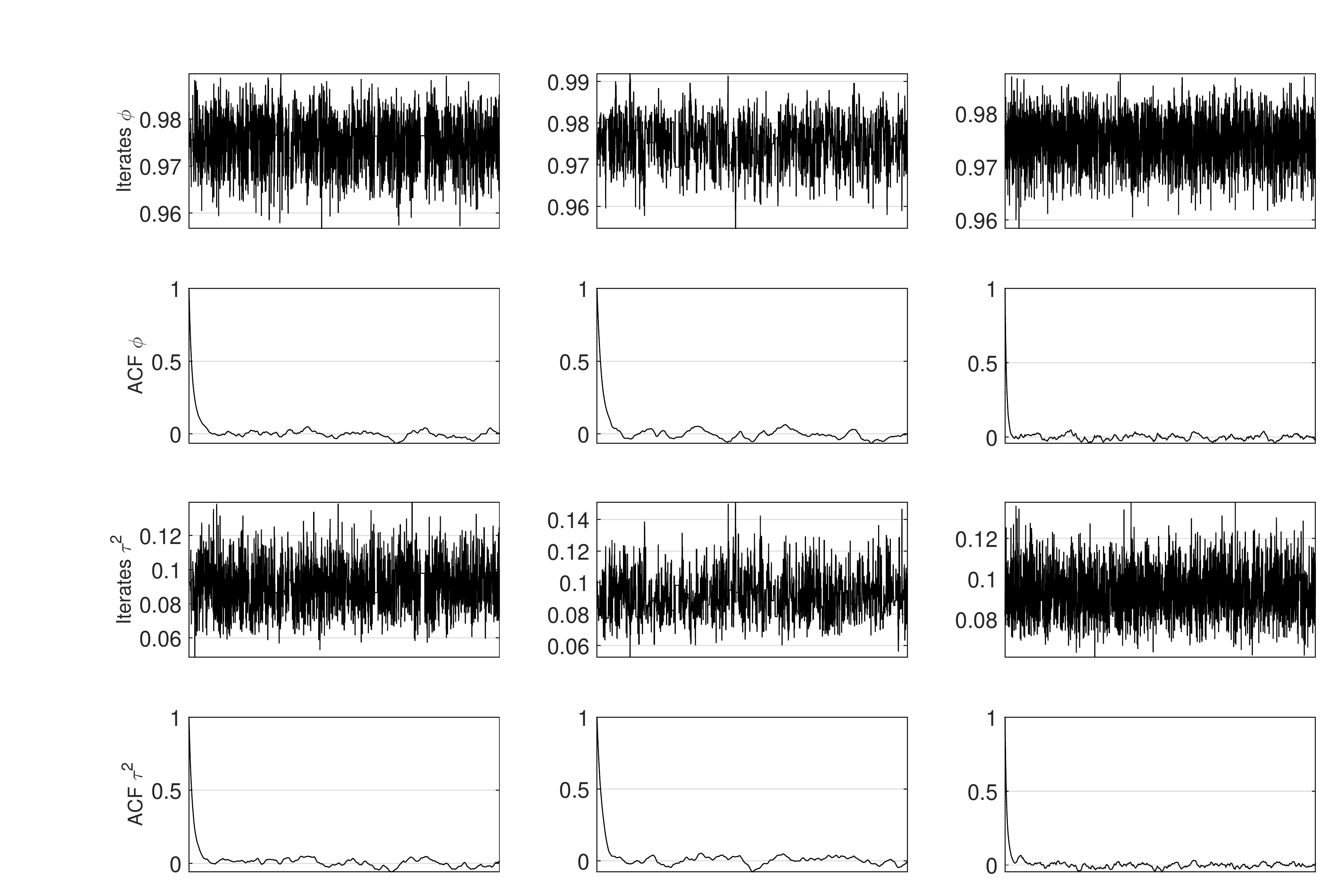}
\vspace*{0.1in}
\caption{The estimation result for the S\&P $500$ daily financial returns data applied to the AR($4$) SV model corresponding to the result reported in Table~\ref{tab:sv:finance3564}. The columns from left to right correspond to, (a) the CPM with sorting the univariate disturbances, (b) the BPM with RW, and (c) the BPM with the proposal using the derivative information.}
\label{fig:sim:sv4:7832}
\end{figure*}
\end{appendices}

\end{document}